\documentclass[%
 rsi,
 amsmath,amssymb,
 reprint,%
]{revtex4-1}

\usepackage{dcolumn}

\usepackage[utf8]{inputenc}
\usepackage[T1]{fontenc}
\usepackage{mathptmx}

\usepackage{graphicx}  
\usepackage{dcolumn}   
\usepackage{bm}        

\usepackage{color} 

\usepackage{comment}
\usepackage{tabularx}
\usepackage{libertine}

\newcommand{\erm}{{\rm e}}
\newcommand{\beq}{\begin{equation}}
\newcommand{\eeq}{\end{equation}}

\begin{document}
\title{Site-selection criteria for the Einstein Telescope}
\author{Florian Amann$^{1,2}$ and Fabio Bonsignorio$^3$ and Tomasz Bulik$^{4}$ and Henk Jan Bulten$^{5,6}$ and Stefano Cuccuru$^{7,8}$ and Alain Dassargues$^{9}$ and Riccardo DeSalvo$^{10,11}$ and Edit Fenyvesi$^{12,13,14}$ and Francesco Fidecaro$^{15,16}$ and Irene Fiori$^{17}$ and Carlo Giunchi$^{18}$ and Aniello Grado$^{19,20}$ and Jan Harms$^{21,22}$ and Soumen Koley$^5$ and L\'aszl\'o Kov\'acs$^{23}$ and Giovanni Losurdo$^{16}$ and Vuk Mandic$^{24}$ and Patrick Meyers$^{25}$ and Luca Naticchioni$^{26,27}$ and Fr\'ed\'eric Nguyen$^{28}$ and Giacomo Oggiano$^{7,8}$ and Marco Olivieri$^{29}$ and Federico Paoletti$^{16}$ and Andrea Paoli$^{17}$ and Wolfango Plastino$^{30,31}$ and Massimiliano Razzano$^{15,16}$ and Paolo Ruggi$^{17}$ and Gilberto Saccorotti$^{18}$ and Alicia M Sintes$^{32}$ and L\'aszl\'o Somlai$^{12,33}$ and Peter V\'an$^{12,34}$ and Matyas Vas\'uth$^{12}$}

\affiliation{$^1$Department of Earth Sciences, ETH Zurich, Zurich, Switzerland}
\affiliation{$^2$Chair of Engineering Geology, RWTH Aachen, Aachen, Germany}
\affiliation{$^3$Heron Robots srl, I-16121 Genova, Italy}
\affiliation{$^4$Astronomical Observatory Warsaw University, 00-478 Warsaw, Poland}
\affiliation{$^5$Nikhef, Science Park 105, 1098 XG Amsterdam, The Netherlands}
\affiliation{$^6$VU University Amsterdam, 1081 HV Amsterdam, The Netherlands}
\affiliation{$^7$Dipartimento di Chimica e Farmacia, Universit\`a degli Studi di Sassari, 07100, Sassari, Italy}
\affiliation{$^8$INFN Laboratori Nazionali del Sud, 95125 Catania, Italy}
\affiliation{$^9$Hydrogeology and Environmental Geology, Urban \& Environmental Engineering (UEE), University of Li\`ege, Belgium}
\affiliation{$^{10}$Riclab LLC, 1650 Casa Grande Street, Pasadena, CA 91104, USA}
\affiliation{$^{11}$University of Sannio at Benevento, Benevento I-82100, Italy}
\affiliation{$^{12}$Wigner Research Centre for Physics, Institute of Particle and Nuclear Physics, 1121 Budapest, Konkoly Thege Mikl\'os út 29-33}
\affiliation{$^{13}$Institute for Nuclear Research (Atomki), Bem t\'er 18/c, H-4026 Debrecen, Hungary}
\affiliation{$^{14}$University of Debrecen, Doctoral School of Physics, Bem t\'er 18/b, H-4026 Debrecen, Hungary}
\affiliation{$^{15}$Universit\`a di Pisa, I-56127 Pisa, Italy}
\affiliation{$^{16}$INFN, Sezione di Pisa, I-56127 Pisa, Italy}
\affiliation{$^{17}$European Gravitational Observatory (EGO), I-56021 Cascina, Pisa, Italy}
\affiliation{$^{18}$Istituto Nazionale di Geofisica e Vulcanologia (INGV), Sezione Pisa, Pisa, Italy}
\affiliation{$^{19}$INAF, Osservatorio Astronomico di Capodimonte, I-80131 Napoli, Italy}
\affiliation{$^{20}$INFN, Sezione di Napoli, Complesso Universitario di Monte S.Angelo, I-80126 Napoli, Italy}
\affiliation{$^{21}$Gran Sasso Science Institute (GSSI), I-67100 L'Aquila, Italy}
\affiliation{$^{22}$INFN, Laboratori Nazionali del Gran Sasso, I-67100 Assergi, Italy}
\affiliation{$^{23}$RockStudy Ltd, P\'ecs, Hungary}
\affiliation{$^{24}$University of Minnesota, Minneapolis, MN 55455, USA}
\affiliation{$^{25}$OzGrav, University of Melbourne, Parkville, Victoria 3010, Australia}
\affiliation{$^{26}$Universit\`a di Roma ``La Sapienza'', I-00185 Roma, Italy}
\affiliation{$^{27}$INFN, Sezione di Roma, I-00185 Roma, Italy}
\affiliation{$^{28}$Applied Geophysics, Urban \& Environmental Engineering (UEE), University of Li\`ege, Belgium}
\affiliation{$^{29}$Istituto Nazionale di Geofisica e Vulcanologia (INGV), Sezione Bologna, Bologna, Italy}
\affiliation{$^{30}$Dipartimento di Matematica e Fisica, Universit\`a degli Studi Roma Tre, I-00146 Roma, Italy}
\affiliation{$^{31}$INFN, Sezione di Roma Tre, I-00146 Roma, Italy}
\affiliation{$^{32}$Universitat de les Illes Balears, IAC3---IEEC, E-07122 Palma de Mallorca, Spain}
\affiliation{$^{33}$Institute of Physics Faculty of Sciences, University of P\'ecs, H-7624 P\'ecs, Ifj\'us\'ag str. 6}
\affiliation{$^{34}$Budapest University of Technology and Economics, Faculty of Mechanical Engineering,  Department of Energy Engineering, Budapest, Hungary}

\begin{abstract}
The Einstein Telescope (ET) is a proposed next-generation, underground gravitational-wave (GW) detector to be based in Europe. It will provide about an order of magnitude sensitivity increase with respect to currently operating detectors, and furthermore, extend the observation band towards lower frequencies, i.e., down to about 3\,Hz. One of the first decisions that needs to be made is about the future ET site following an in-depth site characterization. Site evaluation and selection is a complicated process, which takes into account science, financial, political, and socio-economic criteria. In this paper, we provide an overview of the site-selection criteria for ET, provide a formalism to evaluate the direct impact of environmental noise on ET sensitivity, and outline the necessary elements of a site-characterization campaign.
\end{abstract}

\maketitle

\section{Introduction}
The environment surrounding modern fundamental physics experiments assumes an increasingly important role with great impact on infrastructure, cost, and science. In experiments to search for rare particle interactions like the neutrino-less double-beta decay or interactions with dark matter, the local radioactive environment and particle backgrounds can limit the sensitivity of the experiments \cite{ArEA2013a,ApEA2011,AgEA2014,AkEA2015,AlEA2015,AlEA2017}. Modern particle detectors are located underground to reduce the natural background. Sites of new ground-based telescopes have to be chosen carefully to enable excellent seeing conditions and to avoid light pollution \cite{TOEA2010,RiEA2008,VeEA2011,VaEA2012,VaEA2014}. Sometimes, the environment can even form an essential component of the experiment itself like in large-scale neutrino detectors \cite{dJEA2010,AaEA2017}. Even at CERN, where the direct impact of the environment can be corrected by feedback and plays a minor role, environment-dependent aspects of infrastructure lifetime are of great importance and need to be analyzed \cite{dMu2019}. Site characterization and selection is therefore of great value in large modern fundamental-physics experiments and can crucially influence their future scientific output.

The environment plays an even more important role for gravitational-wave (GW) detectors. For the LIGO and Virgo detectors, the site conditions were assessed especially with respect to the feasibility of the construction, but also the importance of having an environment with weak seismic disturbances was emphasized \cite{LIGO1989,AcEA2012}. Ground motion, sound, and other environmental noises can directly affect the sensitivity and duty cycle of a GW detector \cite{EfEA2015}. For Einstein Telescope (ET), general site conditions concerning, for example, geology and ground water can have a great impact on construction cost, infrastructure lifetime, and environmental noise. A preliminary seismic assessment of numerous sites in Europe was carried out as part of the ET Conceptional Design Study \cite{ET2011,BBR2015}. One of the goals of ET is to extend the frequency band of ground-based GW observations down to a few Hertz \cite{HiEA2011}, which amplifies the importance of environmental noise. Seismic fields were given special attention since the main environmental noise predicted to set a low-frequency limit to ET's bandwidth is from gravity perturbations produced by seismic fields \cite{BaHa2019,Har2019}. Among the environmental noises, terrestrial gravity perturbations, if they limit the detector sensitivity, require a complicated mitigation method \cite{Har2019}. Suppressing terrestrial gravity perturbations is the main motivation to construct ET underground and therefore determines a large fraction of the cost.

Two candidate sites were chosen to be subject to a detailed site-characterization: north of Lula in Sardinia (Italy), and the Meuse-Rhine Euroregion. It is the responsibility of the ET collaboration to present an evaluation of the two sites. A site evaluation needs to consider the impact of site conditions on:
\begin{itemize}
\setlength\itemsep{0em}
    \item Detector sensitivity
    \item Detector operation and duty cycle
    \item Infrastructure lifetime
    \item Site-quality preservation
    \item Construction and maintenance cost
    \item Socio-economic impact of ET
\end{itemize}
Individual environmental properties such as local geology, topography, and seismic activity can be relevant to more than one of these criteria. While it is helpful to introduce these categories for a detailed discussion, the ultimate question is what the achievable quality of a detector is in terms of sensitivity, duty cycle, and its socio-economic impact integrated over the lifetime of the infrastructure for a given amount of invested money. There is no algorithm nor theory to fully answer the question, but discussions leading to a site selection must be oriented towards an answer to this simply stated problem. 

The goal of this paper is to help prioritizing the criteria and to facilitate the site selection. A complete description of site-selection criteria as quantitative measures, for example, to estimate cost and social impact is well beyond the scope of this article. Instead, we provide a summary of the respective site properties that will have to be studied for site selection. We limit the quantitative analysis to aspects that have a \emph{direct} impact on ET's sensitivity, i.e., the calculation of environmental noise, neglecting relations that exist between all criteria due to financial constraints.

In section \ref{sec:siteconditions}, we discuss general site conditions related to geology, ground water, etc. In section \ref{sec:envnoise}, we describe environmental noises and how to estimate associated ET instrument noise. Since site characterization plays such an important role, we summarize the targets of a site-characterization campaign in section \ref{sec:sitechar} and how to obtain the required information.

\section{Site conditions}
\label{sec:siteconditions}
In this section, we discuss the site-selection criteria from an infrastructural and geological point of view. This should include all the possible parameters that have an impact on the excavation costs and construction timeline, detector operation, underground facility access convenience, safety of the workers in the underground environment and detector lifetime that we assume to be at least 50 years. The parameters related to the underground facilities have been grouped in terms of geological conditions, hydrogeological conditions, and geotechnical conditions. Another section concerns surface conditions, infrastructures, and societal aspects. 

The main goal of site selection, site characterization, facility layout, and identification of applied construction methods is to find a location that allows for the construction of ET so that it can achieve its science goals and operate effectively for its proposed lifetime. The technical and cost aspects, nevertheless, can only be optimized together, as a result of a multi-component decision-making procedure, balancing among sensitivity, cost and technical risk analyses. The most reasonable solution for the selected site, the basic design and the planned construction methods should ensure optimization both for technical readiness and the overall costs (both for construction and operation phases) of the facility. 

\subsection{Geological conditions}
The challenges related to the construction of a deep (down to 300\,m) and long (more than 30\,km of total tunnel length and experimental halls) infrastructure such as ET are many and most of them are related to the difficulty to anticipate the geological conditions  (structures, faults, lithology, fractures, alteration, short- and long-term water ingress, ...) at depths and their corresponding hazards over large scales \cite{ZYL2019}. Construction planning needs to consider structural information, geological, rock mechanical and behavior models including maps and cross-sections with an estimate of uncertainties, and to estimate risks of geological hazards (i.e. tunnel stability, environmental impact such as ground water lowering and subsidence, karst, earthquakes). This process should use the most advanced combination of methods to predict the geological and rock-mechanical conditions \cite{ChEA2011b}, and the impact underground construction will have on it. For example, changes in groundwater conditions have been induced by underground construction.

Seismicity plays a large role in the duty factor of large ground-based, gravitational-wave experiments \cite{MuEA2019a}. Specific aspects of geology in relation to seismicity are site effects and seismic microzonation \cite{MiEA2011}. We can have variations of seismic amplitude at small scales due to filtering, attenuation, and amplification \cite{SeEA2005,DoEA2009}. Filtering is the frequency-dependent transmission of seismic waves, for example, through stratified geology. Amplification under “stable conditions” is the effect of the interference of seismic waves trapped within geological bodies bounded by large seismic impedance contrasts (soft soil/bedrock, soil/free surface, etc.). The dimension of geological bodies and discontinuities to be analyzed for characterizing the relevant phenomena are of the order of the seismic wavelengths, which can range from several tens of meters to several kilometers depending on frequency and ground properties. The rate of attenuation, typically expressed as the attenuation factor Q, depends on a variety of ground properties such as the elastic properties, degree of fracturing, presence of ground water, fluid pressure and porosity. While the impact of site effects is straight-forward to understand with respect to ground vibration, and therefore to detector control and seismic isolation, a more detailed understanding of the geology leading to site effects would be required for models of seismic terrestrial gravity noise (see section \ref{sec:seismic}). 
 
\subsection{Hydrogeological conditions}
Hydrogeological conditions govern the groundwater flow. Water inflow into tunnels, shafts and larger cavities is an important factor during the construction phase as well during the exploitation phase. Depending on the permeability, the accumulated water inflow rates can be high requiring a tunnel drainage system designed for pumping water back to the surface \cite{Hem2012}. Pumping is associated with ambient noise and the source of noise is at the depth of ET. Zones where large short and long-term inflow rates are expected might be treated with cement injections to decrease their permeability and thus reduce the accumulated water ingress significantly. If a water-drainage system employs pumps, they might be a significant source of infrastructure noise affecting the GW detector. Water flow as part of a drainage system inside tunnels might potentially act as a source of gravity noise \cite{AkEA2018a}. Groundwater constitutes a possible hazard scenario for deep infrastructures \cite{Col2014}.

Hydrogeological data must be collected at the relevant scale (typically meter to decameters) corresponding to the different lithological facies (i.e., nature of the geological formations) that can be potentially encountered. Hydrogeological data can be provided at the intact rock scale (a rock specimen that does not contain any fractures/joints) and on the rock-mass scale (a volume of jointed rock). Since flow in the underground is often controlled by flow in fractures, the permeability is typically higher on the rock-mass scale. Hydraulic conductivity and storativity as well as water pressures or piezometric heads, are the most important parameters and variables determining the quantity of groundwater to be potentially drained by underground galleries and cavities. As hydraulic conductivity in a rock mass is highly dependent on faulting, local degree of fracturing, fracture connectivity and fracture apertures must be considered. Another big issue is certainly the depth-dependent values for hydraulic conductivity in a given lithology. Due to potentially depth-dependent hydraulic data, it is required to obtain these data from packer tests along the trajectory of  wellbores down to the target depth of ET. Tunneling induced, transient pore-pressure changes cause a poro-elastic effect in the reservoir and may lead to surface subsidence. In karstic limestones, the hydrogeological parameters are quite heterogeneous, with the hydraulic conductivity varying locally by several orders of magnitude leading to a poor ‘representativity’ of most of the field and borehole in situ tests and measurements. For any hydrogeological context, the acquired values from future field tests would need to be processed with care, and conservative assumptions would be needed for all future hydraulic and stability calculations. Variables and parameters to account for are \cite{GuWa2012,Hol2014,Das2018}:
\begin{itemize}
\setlength\itemsep{0em}
\item Water quantity and quality variables:
\begin{itemize}
\item Water pressures / piezometric heads 
\item Solutes concentrations (hydrochemistry)
\end{itemize}
\item Hydrogeological parameters:
\begin{itemize}
\item Hydraulic conductivity
\item Porosity
\item Storativity
\item Effective drainage porosity
\end{itemize}
\end{itemize}
All those data are to be integrated in:
\begin{itemize}
\item Hydrogeological models:
\begin{itemize}
\item Hydrogeological maps and cross-sections
\item 3D conceptual model of groundwater flow
\end{itemize}
\item Hydrogeological hazard assessment:
\begin{itemize}
\item In the construction phase (transient)
\item In the exploitation phase (assumed steady state)
\end{itemize}
\end{itemize}

\subsection{Geotechnical conditions and infrastructure}
The general aim of rock-mechanical data acquisition is to understand and forecast the behavior of the host rock mass and the variability of the parameters/processes/phenomena as a function of rock types, weathering level, parting, lateral and vertical position, anisotropy, etc. This makes it possible to develop a robust hazard catalogue for risk assessment and counter-measure design, to reduce the uncertainties in rock mechanical data for static calculations, and to ensure the technical/economic optimization of the facility. In addition, rock mechanical parameters have an influence on seismic noise, specifically its attenuation from the surface to the underground location of ET. Some of the important geomechanical parameters and features to consider include:
\begin{itemize}
\setlength\itemsep{0em}
\item Faults and fractures
\item Rock mechanical data:
\begin{itemize}
\item Elastic parameters (static and dynamic Young's modulus and Poisson's ratio of the intact rock, and the rock mass)
\item Strength parameters (uniaxial and triaxial compressive strength, tensile strength, shear strength of intact rocks and discontinuities)
\end{itemize}
\item In-situ stresses
\item Rock-mass characterization
\item Geomechanical hazards, e.g., squeezing, wedge failure, unravelling, face stability, swelling, subsidence, and other hazards related to the excavation method
\end{itemize}
The detailed design of the ET infrastructure will be based on rock-mass characterization, which includes the spatial distribution of rock-mass types along the ET alignment, stress information, excavation method, excavation geometry and related hazard scenarios.   

Radioactivity is to be considered for the safety of the workers at underground sites \cite{BMG2016}. The primary radioactive elements in the Earth’s crust that leads to human exposure are potassium, uranium, thorium, and their radioactive decay products (e.g. radium, radon) \cite{OjLe2014}. The majority of the dose to the lung arises from exposure to the short-lived decay products of radon and thoron. Radon and thoron are ubiquitous in the air at ground level and are significant contributors to the average dose from natural background sources of radiation. In homes, in underground mines and in other situations where radon (and thoron) may be present and where ventilation may be limited, the levels of these radionuclides and their decay products can accumulate to unacceptably high levels. Soils and rocks are often the main sources of radon. In unsaturated soils or rocks, radon moves in gaseous form through pores and fractures. In saturated zones, radon moves in solution into groundwater to underground openings, such as mines and caves, and to buildings. For underground facilities it is important to consider the contribution from the outdoor environment through the ventilation system and from building materials. While most building materials produce small amounts of radon, certain materials can act as significant sources of indoor radon. Such materials have a combination of elevated levels of $^{226}$Ra (the radioactive parent of radon) and a porosity that allows the radon gas to escape. Examples are lightweight concrete with alum shale, phosphogypsum and Italian tuff. EURATOM establishes reference levels for indoor radon concentrations and for indoor gamma radiation emitted from building materials. Recent epidemiological findings from residential studies demonstrate a statistically significant increase of lung cancer risk from prolonged exposure to indoor radon at levels of the order of 100\,Bq/m$^3$ \cite{YoEA2016}. 

\subsection{Surface infrastructure and societal aspects}

Even though ET's main infrastructure will lie underground, surface conditions are very important to the project. Parts of the infrastructure will be located at the surface, including operations buildings, underground access, potentially a visitor center and guest houses. Seismic disturbances created by regional infrastructure, e.g., traffic and industry, can still interfere with the operation of the detector and produce sensitivity limitations. The excavation of caverns and tunnels will produce a large amount of waste rock, which needs to be disposed. In summary, important surface site criteria affecting detector construction are 
\begin{itemize}
\setlength\itemsep{0em}
\item	Main and secondary road and railway networks and their typical load
\item	Existing utilities and technological networks in the area (power, gas, data, water supply, sewage systems)
\item	Presence and classification of wells and water uptake systems
\item	Site availability and acquisition costs
\item   Constraints on the surface access locations to the underground infrastructure, which must also consider safety access along arms
\item	Environmental restrictions (waste control especially with respect to rock disposal, water control, soil conservation, nature and landscape conservation, environmental impact)
\item	Legal issues must be considered for what concerns the authorization procedures and the analysis of territorial constraints.
\end{itemize}
For the support infrastructures, we identify the following parameters:
\begin{itemize}
\setlength\itemsep{0em}
\item	Site accessibility 
\item	Accommodations for resident staff (housing, schools, shopping, etc.) 
\item	Accommodations for visiting staff (hotels, transportation, etc.)
\item	Local technical support (qualified vendors, maintenance, fabrication, etc.)
\item	Site utilities installation (power, water, etc.)
\end{itemize}
Surface parameters that are important to detector operation:
\begin{itemize}
\setlength\itemsep{0em}
\item	Climate and environmental risks (earthquakes, floods, wind speeds, precipitation, lightning rate)
\item	Cost of power
\item	Heating and cooling requirements of underground caverns (in combination with humidity control)
\item	Maintenance requirements
\item	Travel time and costs for visiting staff
\item	Cost and quality of living
\end{itemize}

In addition, societal and economic considerations reported in a socio-economic impact assessment can lead to important distinctions between sites. While a comprehensive discussion of the relevant aspects of these assessments is beyond the scope of this paper \cite{BMEA2004,Gov2008}, certain aspects are directly relevant to the involved scientists. Most importantly, the relation between the local population and a scientific project can be crucial for the realization of a project. The spread of misinformation and the disregard of local interests has led to construction delays or even shut-down of experiments in the past \cite{Nos2003,Wit2019,KaEA2020}. Early outreach activities before the start of construction help to correctly inform local people and to understand the relation of the local population to the planned experiment, and thereby give the possibility to address issues before final decisions about the construction plan are taken. 

\subsection{Infrastructure lifetime and cost factors}
\label{sec:lifecost}
\subsubsection{Tunneling costs}

Tunneling differs from the construction of other infrastructure in many ways. The main issues that distinguish tunnels from other infrastructure arise from the risk involved with excavation through unknown ground conditions and the numerous individual cost drivers that  contribute to the overall cost. These cost drivers include, but are not limited to the following direct and indirect factors,

\begin{itemize}
\setlength\itemsep{0em}
\item Excavation volume (i.e., tunnel length and diameter)
\item Ground conditions and related uncertainties
\item Ground behavior
\item Excavation method
\item Tunnel depth
\item Support requirements
\item Final lining design
\item Water ingress and tunnel drainage system
\item Environmental aspects
\item Labour cost
\item Health and safety regulations 
\item Market competition
\item Government and public support 
\item Contract type 
\item Cost of bidding
\end{itemize}

Geology can range from soft to hard rocks and can include shear zones. A site investigation must be completed during the initial design stages of a project to account for and plan for various ground conditions, and to estimate costs. Varying geologies necessitate different methods of excavation, which include drill and blast, roadheaders, and tunnel boring machines (TBMs). In addition to all of these variables, tunneling is also affected by many indirect factors often related to the country of construction as each differs in its labour costs, health and safety regulations, environmental regulations, level of market competition, client knowledge, and amount of government and public support. Varying contract types such as design and construct (D\&C); design, build, operate (DBO); build, own, operate (BOO); and public private partnerships (PPP) are also common in different countries and affect the cost of bidding and financing. It should also be mentioned that excavation cost for ET can be greatly reduced if topography of a site makes it possible to have most of the vacuum pipes above ground. Only the test masses are required to be located sufficiently deep underground.

\subsubsection{Lifetime}

The ET infrastructure should have a lifetime greater than 50 years. Parameters to be considered in this respect concern stability and corrosion:
\begin{itemize}
\setlength\itemsep{0em}
\item Differential deformations within the rock mass including dislocation on active faults or subsidence across each of the 10\,km arms need to be sufficiently small. Requirements need to be set across short distances (the extent of vacuum pipe modules) to limit stress on welding lips (a few mm of differential motion per 15\,m segment is the limit for Virgo), and across long distances to constrain the position of the optical axis. 
\item Atmospheric corrosion is influenced by average and peak humidity in the caverns and tunnels, the pH of ground and condensation water, and by the presence of chemical elements (in particular chloride if stainless steel will be used for the pipes) \cite{LiLi1999}
\item Microbiologically Influenced Corrosion \cite{Lew2009}
\item AC-induced corrosion due to nearby high voltage electric power lines \cite{TrMe2014}.
\end{itemize}
The preservation of the site quality in terms of environmental seismic disturbances over the entire ET lifetime is also important. Regional environmental seismic noise can increase due to the emergence of new industry and traffic including, for example, wind farms, rail service, industry, and mining. This can impact detector sensitivity and operation. Extensive studies of existing and potential future regional sources of seismic disturbances were carried out for the LIGO, Virgo, and GEO600 detectors \cite{Sch2002,FHP2003,DaEA2004,FiEA2009,SaEA2011}. Hence, the question arises if there are characteristics of a site that make it more likely that site quality can be maintained. Similar studies will also be vital for the ET site selection. In addition, agreements with local authorities, made before site selection, that stipulate a minimum distance between major noise sources and ET are mandatory. The higher the quality of a site, the more effort needs to be done to maintain its quality, but one can expect that noise-exclusion areas are easier to obtain in less populated regions. 

\section{Environmental noise model}
\label{sec:envnoise}

\subsection{Seismic field}
\label{sec:seismic}
Sufficiently strong seismic disturbances can reduce the duty cycle of a detector by causing intermittent failures of the interferometer control systems. Such an event is referred to as a lock loss. The main source of seismic disturbances causing these failures are earthquakes \cite{CoEA2017,MuEA2019a}, but even a strong local, anthropogenic source might cause lock loss. However, since the underground environment and the maintenance of a low-noise area around ET (see section \ref{sec:lifecost}) will provide a certain level of protection from anthropogenic sources, and due to recent progress with providing early warnings of earthquakes to gravitational-wave detectors and with the development of control strategies to counteract the impact of strong ground motion \cite{BiEA2018,MuEA2019}, one might expect that the reduction of the duty cycle of ET by seismic disturbances will be modest. More important is the generation of noise in the detector data by ambient seismic fields. 

Seismic displacement of the Earth's surface or underground can couple to the detector output via different mechanisms. First, seismic ground motion can cause noise in GW data through scattered light, which means from stray light interacting with structures that are weakly isolated or not at all isolated from ground motion \cite{accadia2010noise}, or by directly displacing the test masses due to the residual low-frequency seismic noise that passes through the seismic-isolation system \cite{AcEA2010}. Furthermore, seismic noise complicates the controls of the seismic filter chain, giving rise to additional control noise \cite{DoEA2013,Mar2015,MLMa2019}. Last, the seismic displacement and density fluctuations of the ground medium due to seismic-wave propagation can couple to the test masses through gravitational forces and introduce noise in the GW data. This noise is referred to as Newtonian noise (NN) or gravity-gradient noise \cite{Sau1984}.

Seismic fields can be described as solutions to the elastic equation of a medium \cite{AkRi2009}. This equation can under certain assumptions be cast into the form of wave equations, and solutions to these equations traveling through Earth are known as body waves. Based on the particle motion and the direction of propagation of the body waves, they can be categorized into P-waves (compressional waves) and S-waves (shear waves). However, when the medium is bounded, other wave types are generated, which travel along the surface of the medium and are known as surface waves. Depending on the polarization of the particle displacement they can be categorized into Rayleigh and Love waves. Unlike Rayleigh waves, Love waves cannot exist in a homogeneous half-space and require a layered geology.

Seismic displacement is a combination of both body and surface waves. The ratio between the body-wave and the surface-wave content essentially depends on the type of sources (point or line sources), location of sources (surface or underground), damping coefficient of the propagation medium (intrinsic attenuation) and the distance of the observation point from the source \cite{kim2000propagation}. In a homogeneous half-space, amplitudes of body waves decay as $1/r$ in the interior of the medium and with $1/r^2$ at the surface, and surface waves decay with $1/\sqrt{r}$ in the far field of sources, where $r$ is the distance from the source. Hence, considering only geometric attenuation, body-wave amplitude decays faster as compared to surface waves when moving away from the source. However, the intrinsic attenuation of wave amplitudes is a frequency-dependent phenomenon and expressed as $\exp(-\pi f x/ (Q v))$ where $v$ represents the wave velocity at frequency $f$, $x$ the propagation distance, and $Q$ represents the quality factor of the medium \cite{WaSa1996}. Consequently, in a multilayered medium where surface-wave dispersion is observed \cite{haskell1953dispersion}, high-frequency surface waves with wavelengths much shorter than body waves undergo larger attenuation than body waves. Hence, what type of wave dominates surface displacement depends not only on source characteristics, but also crucially on the distance to the sources \cite{HaOR2011,CoEA2018b}.

\begin{figure}[ht!]
\includegraphics[width=0.9\columnwidth]{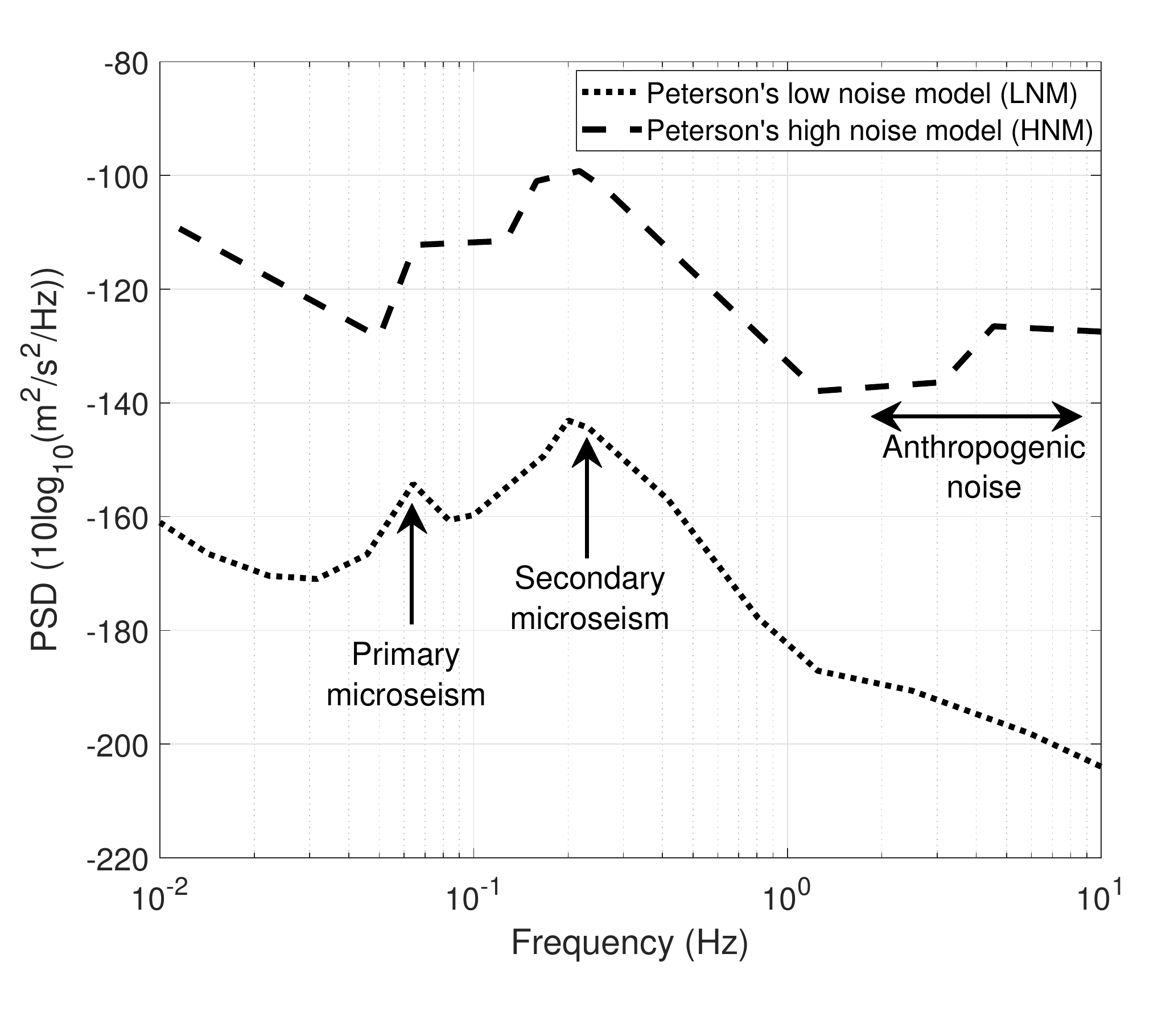}
\caption{Power spectral density (PSD) of Peterson's high noise and low noise models derived from worldwide observations. The models approximately set the lower and the higher limit to globally observed seismic noise PSDs.}
\label{fig:Peterson}
\end{figure}

The Einstein Telescope design sensitivity (see below, figure \ref{fig:envnoise}) is expected to be susceptible to NN below a few tens of Hz. Seismic noise sources active in this frequency band are both natural and anthoprogenic in origin. Anthropogenic sources include traffic (trains and cars), and local human activities, whereas common natural sources are fault ruptures, atmospheric pressure fluctuations, wind interacting with the surface, and ocean waves. The global ambient seismic noise comprising of high and low noise models are shown in figure \ref{fig:Peterson} based on studies by Peterson, 1993 \cite{Pet1993}. Primary microseisms in the frequency band below 0.1\,Hz due to interaction of ocean waves with sea floor are attributed mostly to activation in shallow sea \cite{haubrich1969microseisms}. In the frequency band 0.1 to 0.4\,Hz, the secondary microseisms dominate the noise spectrum. They occur at twice the frequency of ocean waves originating from the non-linear interaction of standing ocean waves causing a pressure wave propagating towards the ocean floor \cite{LH1950}. As shown in figure \ref{fig:Peterson}, a falling seismic-noise amplitude is observed from 0.5 to 1.5\,Hz. An increase in noise amplitude in this band is observed during storms or other extreme meteorological conditions.

\begin{figure}[ht!]
\includegraphics[width=0.9\columnwidth]{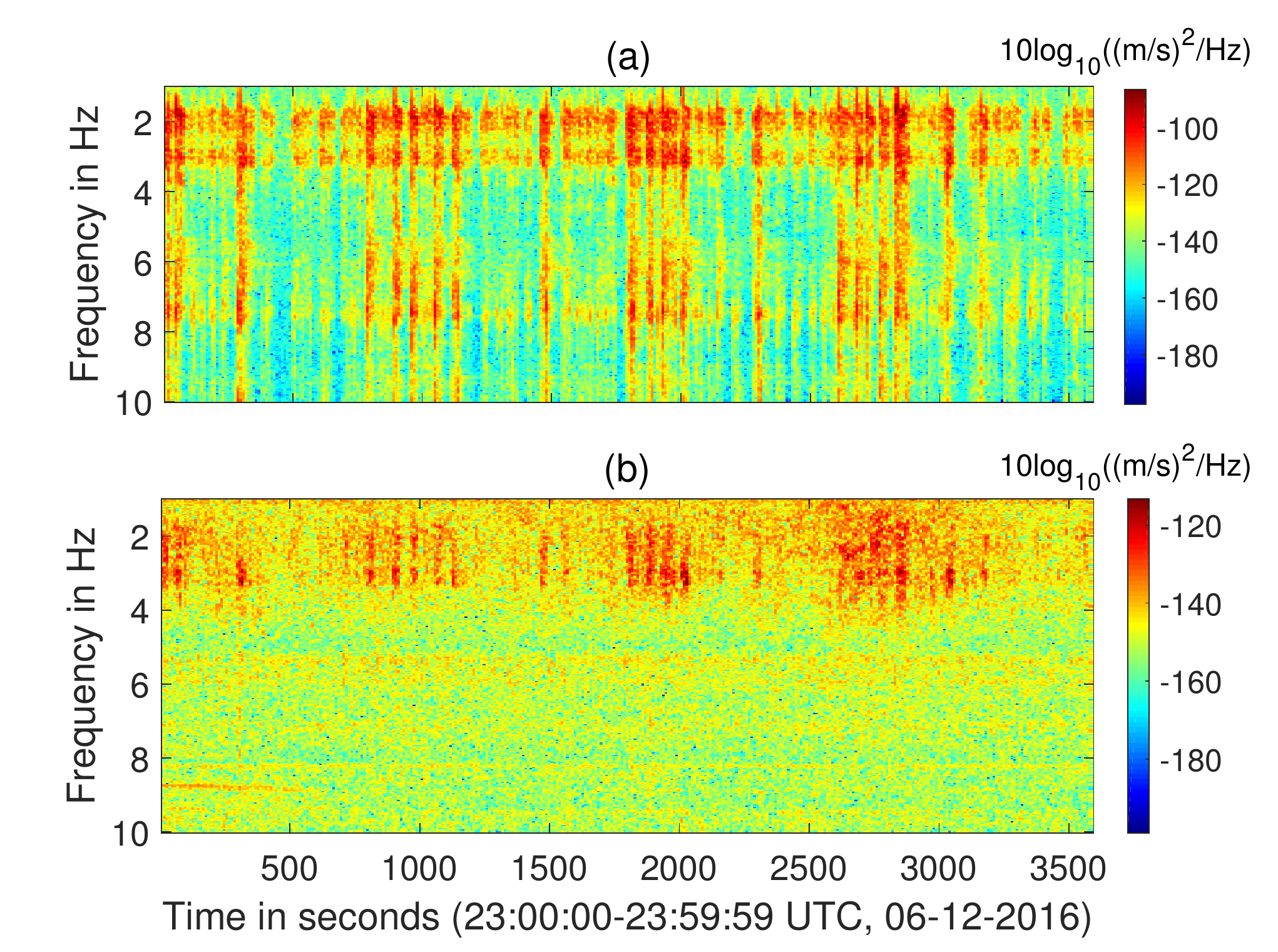}
\caption{(a) Spectrogram of seismic ground velocity measured underneath a bridge 1.5\,km away from the Virgo Central Building. (b) Spectrogram of seismic ground velocity measured at the Virgo Central Building during the same period as in (a). Seismic noise below 4\,Hz is observed to be well correlated between the two sites.}
\label{fig:VirgoNoise}
\end{figure}

At frequencies greater than 1.5\,Hz, seismic noise originating from human activities contributes significantly. This includes noise originating from roads, bridges, industries and use of machinery near the site. Figure \ref{fig:VirgoNoise}(a) and (b) show the spectrograms of the ground velocity measured underneath a bridge (1.5\,km away from the Virgo Central Building) and at the Virgo Central Building (CEB), respectively. In the frequency band 2 to 4\,Hz, imprints of the ground velocity measured underneath the bridge are observed in the measurements at the Virgo CEB \cite{koley2017s}. In the high frequency band above 5\,Hz, local sources at the detector site start to contribute leading to transients from human activity, and also several high-frequency stationary sources of noise like air conditioners, chillers, and mechanical vacuum pumps (e.g., turbomolecular pumps and scroll pumps), which are used for operation of a GW detector, are important on-site sources of noise and must be accounted for while computing the associated NN.

The seismic-noise budget for ET (presented below in figure \ref{fig:envnoise}) includes mechanical coupling through the isolation system using a model developed for ET's Conceptual Design Study \cite{ET2011}, and NN from surface and body waves. The seismic-noise model requires an estimate of underground seismic displacement and ground tilt. Underground seismic displacement is modeled as a sum of two components: surface displacement assumed to be dominated by Rayleigh waves attenuated with depth, and body-wave displacement. The attenuation of Rayleigh-wave displacement with depth is calculated using a dispersion curve of Rayleigh waves. Here, we model it as
\beq
c(f)=2000\,{\rm m/s}\cdot\erm^{-f/4\,\rm Hz}+300\,{\rm m/s},
\eeq
While this model does not represent a specific site, it yields realistic values for the frequency range 1\,Hz to 100\,Hz, which might well be representative of some site \cite{BoEA2002ch2}. Estimation of underground displacement from Rayleigh waves is based on equations that can be found, for example, in \cite{HaNa1998}. The body-wave seismic spectrum is assumed to be independent of depth. This is not strictly guaranteed since reflection of body-waves from the surface can cause depth-dependent amplitudes, and seismic amplitudes can always vary strongly in the vicinity of dominant nearby sources, but whenever the body-wave field is composed of many waves at all frequencies from distant sources, then the assumption of a depth-independent spectrum should be at least approximately valid. Our fiducial seismic spectra used for noise projections in Figure \ref{fig:envnoise} correspond to 5 times the New Low-Noise Model (NLNM) \cite{Pet1993} for the body-wave spectrum, and the logarithmic average of the NLNM and New High-Noise Model for the Rayleigh-wave vertical surface-displacement spectrum. The logarithmic average produces a spectrum that lies in the middle between the low-noise and high-noise models when plotted with logarithmic scale, which is representative of the noise at a typical remote surface site. The tilt spectrum can be estimated from the displacement spectra by multiplication with $2\pi f/v(f)$, where $v$ is the speed of Rayleigh or body waves. Note that this method would underestimate ground tilt at the surface where direct forcing of objects and atmosphere can produce large tilts in addition to the tilt associated with seismic waves \cite{DABo2012}, but it is approximately valid underground.

The underground seismic displacement and tilt spectra are passed through a model of a 17\,m isolation system (similar in design to the Virgo Superattenuator \cite{BrEA2005}). Here, we assume that ground tilt, and horizontal and vertical displacements are uncorrelated, but this is mostly to simplify the calculation and has a minor impact on the seismic noise in ET. Finally, it is assumed that seismic noise entering through different test masses is uncorrelated above 3\,Hz. This should reflect the real situation since seismic waves at 3\,Hz have at most a length of 1 -- 2\,km, while the separation of test masses is 10\,km \cite{PLB2009}. As a caveat, the triangular configuration of ET might lead to some correlation of environmental noise between test masses of \emph{different} interferometers. While this does not influence the noise model, it might well be an important fact for GW data analysis.

When estimating NN for ET, it is again important to consider contributions from Rayleigh waves and body waves. Here, one also needs to know what the relative contribution of shear and compressional waves to the body-wave field is. We assume that $p\equiv S_{\rm P}(\xi_x;f)/S_{\rm bw}(\xi_x;f)=1/3$ of the seismic spectral density from body waves is produced by compressional waves (P waves), where $\xi_x$ is the horizontal displacement along the arm, i.e., we assume that all three body-wave polarizations carry the same average displacement power. Furthermore, it is assumed that the body-wave and Rayleigh-wave fields are (3D and 2D) isotropic. This is certainly an invalid approximation, but it would be misleading to assume any specific form of anisotropy, since anisotropy will be different at different sites, different for each vertex of the detector, and different for each wave type. Anisotropies have a significant impact on NN spectra, and how they enter the NN estimate also depends on the details of the model \cite{Har2019}. In principle, seismic NN can be low in one of the three detectors forming the ET triangle if all seismic waves near the vertex travel in a direction right between the directions of its two arms, and perpendicular to the arms at their ends. However, this still leads to seismic NN in the other two detectors, and since it requires plane-wave propagation, sources of these waves must be sufficiently distant, and it is highly unlikely that all relevant distant sources line up in this way. The NN estimate calculated for a highly anisotropic field at one of the LIGO sites lies within a factor 1.5 of the isotropic model at all frequencies \cite{HaEA2020}.

Rayleigh waves produce NN through rock compression, cavern-wall displacement, and through surface displacement. All three effects are added coherently using equations (36), (62), and (94) in \cite{Har2019}. This leads to the following strain spectral density:
\beq
S_{\rm R}^h(f) = (2\pi/ \sqrt{2}\gamma G\rho_{0,\rm surf})^2\mathcal R(f)S(\xi_{\rm v};f)\frac{4}{L^2(2\pi f)^4}.
\eeq
Here, $S(\xi_{\rm v};f)$ is power spectral density of vertical surface displacement from Rayleigh waves, $\gamma$ a parameter with values in the range 0.5--1 quantifying the partial cancellation of NN from surface displacement and compression of the soil by Rayleigh waves, $\rho_{0,\rm surf}$ is the mass density of the surface medium, $L$ the length of ET's detector arms, and $\mathcal R(f)$ describes the NN reduction as a function of detector depth $h$:
\begin{eqnarray}
    r_0(f) &= k_{\rm R}(f) (1 - \zeta(f))\\ \nonumber
    s_h(f) &= -k_{\rm R}(f) (1 + \zeta(f)) \exp(-k_{\rm R}(f)  h)\\
    b_h(f) &= \frac{2}{3} \big(2k_{\rm R}(f) \exp(-q_{\rm P}(f) h) \\
    &\qquad + \zeta(f) q_{\rm S}(f) \exp(-q_{\rm S}(f)  h)\big)\\
    \mathcal R(f) &= |(s_h(f) + b_h(f)) / r_0(f)|^2
\end{eqnarray}
where $k_{\rm R}$ is the wave number of Rayleigh waves, $q_{\rm P}(f) = 2\pi f \sqrt{1 / v_{\rm R}^2(f) - 1 / v_{\rm P}^2(f)}$, $q_{\rm S} = 2\pi f \sqrt{1 / v_{\rm R}^2(f) - 1 / v_{\rm S}^2(f)}$, and $\zeta(f) = \sqrt{q_{\rm P}(f) / q_{\rm S}(f)}$. Here, it is crucial to use an accurate dispersion model $v_{\rm R}(f)$ for the Rayleigh waves since it has an important impact on how NN decreases with increasing depth $h$. Compressional and shear-wave speeds $v_{\rm P},\,v_{\rm S}$, if not provided independently, must be adapted to the Rayleigh-wave dispersion using estimates of the Poisson's ratio or making ad hoc assumptions of the ratio between Rayleigh-, shear-, and compressional-wave speeds. This is necessary since the Rayleigh waves sample rock at varying depth depending on frequency with different effective shear- and compressional-wave speeds of the sampled rock mass (unless the ground is homogeneous). Note that the limit $h\rightarrow 0$ does not mean $\mathcal R(f)\rightarrow 1$ since the contribution from cavern walls must be subtracted from the underground contribution $b_h(f)$ to get a meaningful surface limit (which means to remove the factor 2/3 and the second term in the brackets).

Body waves produce NN through displacement of cavern walls (shear and compressional waves) and through compression of rock (compressional waves). Both contributions are added coherently using equation (62) in the 2019 version of \cite{Har2019}. The contribution of normal surface displacement by body waves can typically be neglected in the frequency range 3\,Hz--20\,Hz. This can be inferred from seismic observations showing that seismic surface spectra are significantly stronger in this band than underground measurements at the same location (as evidenced by many past observations including studies carried out by the GW community \cite{BBR2015,MaEA2018}). Therefore, seismic NN from normal surface displacement is dominated by Rayleigh waves between 3\,Hz and 20\,Hz. Correlations between shear and compressional waves (and also with Rayleigh waves) are also neglected. Note that simple reflection of body waves from the surface causes scattering into different wave types potentially causing such correlations, but this should have a minor influence on the NN spectral density, which is a long-time average, i.e., averaged over many waves. The body-wave NN spectrum then reads \cite{BaHa2019}
\beq
S_{\rm bw}^h(f) = \left(\frac{4}{3}\pi G \rho_{0,\rm ug}\right)^2(3p+1) S_{\rm bw}(\xi_x;f)\frac{4}{L^2(2\pi f)^4},
\eeq
where $S_{\rm bw}(\xi_x;f)$ is the power spectral density of body-wave displacement along the direction of the arm, and $\rho_{0,\rm ug}$ is the mass density of the rock in the vicinity of the cavern. When evaluating these NN models for a specific site, minor estimation errors are to be expected from simplifying assumptions of soil/rock density including seasonal variations of moisture content.

The main optics of ET would be shielded from seismic noise above 3\,Hz. However, parts of the interferometer that interact with the laser beam and which are not suspended from superattenuators are possible sources of scattered-light noise. Due to the motion of a scatterer, the scattered light adds noise to the GW strain data. Noise from scattered light was for example reported in \cite{accadia2010noise,CaEA2013,EfEA2015}. In the following, we briefly describe the main effects, but we do not include this noise in Figure \ref{fig:envnoise} since it is very hard to foresee how much noise from scattered light will contribute. It is possible though that during much of the ET commissioning process, scattered-light noise will be the main environmental noise.

Overall, GW detectors are designed such that only a tiny fraction of the optical power can introduce noise by scattering. If the scatterer vibrates with a displacement amplitude $\delta x_{\rm sc}(t)$ along the beam direction, then the scattered light's phase changes by
\begin{equation}
\delta\phi_{\rm sc}(t) = \frac{4\pi}{\lambda}\delta x_{\rm sc}(t)
\label{eqn1}
\end{equation}
where $\lambda$ is the laser wavelength. The spectral density of equivalent GW strain noise $S_{\rm sc}^h(f)$ introduced by the scattered light can be obtained as a product of a transfer function $T(f)$ with an effective vibration spectrum (as power-spectral density --- PSD) \cite{CaEA2013}:
\begin{equation}
S_{\rm sc}^h(f) = \left|T(f)\right|^2\cdot{\rm PSD}\left[\frac{\lambda}{4\pi}\sin\left(\frac{4\pi}{\lambda}\delta x_{\rm sc}(t)\right)\right].
\label{eqn2}
\end{equation}
The transfer function describes the optical response of the detector to scattered light entering at a specific location of the detector generally including radiation-pressure coupling. Equation (\ref{eqn2}) can be split into two cases depending on the magnitude of the motion of the scatterer. For small bench motion such that $\delta x_{sc}(t) \ll \frac{\lambda}{4\pi} \approx 10^{-7}$\,m, Eq.~(\ref{eqn2}) linearizes as $S_{sc}^h(f) = |T(f)|^2S(\delta x_{\rm sc};f)$. However, for larger bench motion ($\delta x_{\rm sc}(t)\geq10^{-7}$\,m), the induced strain noise $S_{\rm sc}^h(f)$ follows Eq.~(\ref{eqn2}) and is nonlinear in the vibration amplitude $\delta x_{\rm sc}$. This is typically observed at frequencies between 10 and 20\,Hz due to near-field influence of the mechanical sources of noise or due to microseismic activity at frequencies $<$ 1\,Hz. Although the microseismic activity is not in the detection band, its effect can be visible due to up-conversion \cite{OFW2012}. As for the Advanced Virgo interferometer, there were several instances of scattered light noise in its observation band, which were identified and mitigated, and it is expected to remain an important issue at low frequencies in the future. In most cases, sources of noise were devices like cooling fans and vacuum pumps operating in proximity of back-scattering light spots inside the power-recycling vacuum chamber \cite{ScatteredLightO2}. 

\subsection{Atmospheric fields}
\label{sec:atmospheric}
Atmospheric fields constitute the most complex of all environmental noise sources. This is due to the interaction between surface and atmosphere, and the many different processes that can drive atmospheric perturbations \cite{Hol2004}. The main coupling mechanism of the atmosphere with the detector output is through vibrations that it causes of ground and infrastructure through pressure fluctuations or forcing of surface structure by wind, and by direct gravitational coupling \cite{Cre2008,Har2019}. As the indirect vibrational noise is already discussed in Section \ref{sec:seismic}, we can focus here on the gravitational coupling, which gives rise to so-called atmospheric Newtonian noise (NN).

There are two main types of gravitational coupling. First, acoustic fields produce density perturbations in the form of propagating and standing waves. These perturbations are distinct from any others since they exist even in the absence of wind. The main practical complication in the modeling of acoustic gravitational noise is to procure a sufficiently accurate model of acoustic spatial correlations, which depends on the source distribution and possible acoustic scattering. So far, numerical simulations have only been able to include major geometric constraints like the separation of acoustic fields into outdoor and indoor contributions \cite{FiEA2018}. This is important since the sound level inside LIGO and Virgo buildings (and to be expected as well for the ET caverns) is much higher than the ambient acoustic noise outside. Responsible for the excess noise inside buildings are sources like pumps, ventilation systems, etc. For ET, it will be important to avoid any major acoustic noise below 30\,Hz in its caverns, but some mitigation can be achieved by noise cancellation using microphones \cite{Har2019}. External sources of acoustic noise include transients from thunderstorms and other weather related sources, noise from traffic, planes, and people. Atmospheric sources that have an effect on the detector can be located far from the detector since acoustic waves are known to propagate over long distances in the atmosphere with weak damping of their amplitude.

The acoustic NN model in Figure \ref{fig:envnoise} uses a sound spectrum representative of a remote surface site with a value of $\delta p_{\rm atm}(3\,{\rm Hz})=5.7\cdot10^{-3}\,\rm Pa/\sqrt{Hz}$ and $\delta p_{\rm atm}(10\,{\rm Hz})=1.4\cdot10^{-3}\,\rm Pa/\sqrt{Hz}$ \cite{BBB2005}. The coupling model is calculated separately for two incoherent contributions from the atmosphere and the cavern using the same sound spectrum. The cavern sound spectrum might well be higher if it will not be possible to separate noisy machines from the experimental halls that contain the test masses. Calculating the isotropic average of equation (132) in \cite{Har2019} and subsequently the corresponding strain noise from the perturbation of the gravity potential, one obtains the atmospheric acoustic NN as strain spectral density,
\beq
S_{\rm atm}^h(f)=\left(\frac{2 c_{\rm s} G \rho_0\delta p_{\rm atm}(f)}{p_0\gamma f}\right)^2\mathcal I_{\rm iso}(4\pi fh/c_{\rm s})\frac{4}{L^2(2\pi f)^4},
\eeq
where $c_{\rm s}=340\,$m/s is the speed of sound, $\rho_0$ the mean air density, $p_0$ the mean air pressure, $\gamma=1.4$ air's adiabatic coefficient, $L=10\,$km the length of a detector arm, $h$ the detector depth (assumed to be 300\,m), and $\mathcal I_{\rm iso}(x)$ is the isotropically averaged coupling coefficient:
\beq
\mathcal I_{\rm iso}(x) = \frac{\pi}{4}(L_{-3}(x)-I_1(x)+I_2(x)/x+3L_{-2}(x)/x),
\eeq
where $I_n(\cdot)$ is the modified Bessel function of the first kind, and $L_n(\cdot)$ is the modified Struve function. For $x>1$, it can be numerically problematic to evaluate these functions, but for such values the coupling coefficient can be obtained by using the approximation
\beq
\mathcal I_{\rm iso}(x) \approx 3/x^4.
\eeq
Note that even though the gravity perturbation of every specific sound plane wave decreases exponentially with a function of depth $h$, the isotropic average produces a polynomial suppression for sufficiently large depth. This is because the exponential suppression of NN from a single plane wave with depth is determined by the horizontal wave number \cite{Har2019}, which can be very small depending on the wave's direction of propagation practically leading to very weak suppression for waves at close to normal incidence to the surface.

The second contribution, again assumed to be produced by an isotropic sound field, comes from the cavern. It increases with the cavern radius $R$, and for $R\ll c_{\rm s}/(2\pi f)$. Evaluating the integral in equation (132) of \cite{Har2019} not over a half space, but a spherical volume, the cavern contribution takes the form
\begin{eqnarray}
S_{\rm cav}^h(f)&=\left(\frac{2c_{\rm s} G \rho_0\delta p_{\rm cav}(f)}{p_0\gamma f}\right)^2\frac{1}{3}(1-{\rm sinc}(2\pi fR/c_{\rm s}))^2\\ \nonumber
&\qquad\cdot\frac{4}{L^2(2\pi f)^4}.
\end{eqnarray}
Strictly speaking, this expression is accurate only for a half-spherical cavern shape with the test mass at its center, but it still serves as a useful estimate as one can expect that deviations from spherical ceilings can be accounted for by a suitable redefinition of the parameter $R$, and multiplying by a frequency-independent geometrical factor, which does not change the order of magnitude of the noise. These corrections are likely minor compared with other corrections, e.g., from anisotropy of the sound field. In this paper, we use a cavern radius of 15\,m.

With respect to the acoustic NN model shown in Figure \ref{fig:envnoise}, more realistic estimates will likely be smaller since the isotropic plane-wave field assumed in this model yields relatively large spatial sound correlations. Sound scattering or complex source distributions reduce spatial correlations, and therefore increase suppression of gravitational coupling with distance to the atmosphere. We also note that cancellation of atmospheric acoustic NN is highly challenging. Microphones are subject to wind noise produced by wind-driven turbulence around microphones \cite{Gre2015}, but since air flow will be controlled underground, wind noise will not interfere with the cancellation of cavern acoustic NN. Alternative technologies like LIDAR are not yet sensitive enough to monitor acoustic fields in the ET band. Hence, significant contributions of atmospheric acoustic NN are to be avoided.

The second type of gravitational coupling between atmosphere and test masses is wind driven. In the ET observation band, atmospheric temperature and humidity fields, which are both associated with a corresponding density field, can be considered stationary in the absence of wind (or generally, when using the Lagrangian description of a fluid). However, when wind is present, then advected gradients in the density field appear as fast fluctuations at a fixed point. The gravitational coupling depends on the product $2\pi f d/v$, where $v$ is the wind speed, and $d$ is the (shortest) distance between test mass and the moving air. In the simplest case of smooth airflow, the suppression with distance is exponential, i.e., the coupling contains the factor $\exp(-2\pi f d/v)$ \cite{Cre2008}, which means that any form of wind-driven coupling is negligible in ET with $d$ being a few 100\,m. When vortices form around surface structures, then the suppression with distance would not be exponential anymore, but it can still be argued that coupling remains negligible in ET \cite{Cre2015}. Therefore, we have neglected the wind-driven gravitational noise in Figure \ref{fig:envnoise}, but for relatively shallow detector depth of 100\,m or less, it might become important, and advection noise should be included. Cancellation of advection NN is conceivable. This is because the density perturbations associated with temperature and humidity fields are large compared to the density perturbations associated with sound. LIDAR can in fact produce three dimensional tomography of temperature and humidity fields \cite{Beh2005,HaEA2015b,SpEA2016}. In addition, Doppler LIDAR can provide three-dimensional scans of the velocity field \cite{CLN2004}. This information combined is all that is required to estimate and subtract the associated NN.

\subsection{Electromagnetic field}
\label{sec:electromagnetic}
Electromagnetic (EM) disturbances can be produced in many ways including natural sources and self-inflicted noise from electronics \cite{EfEA2015}. The latter includes cross-coupling between electronic/magnetic components of the detector like connectors, cables, coils, and permanent magnets, transients from overhead power lines, and noise from the mains power supply (50\,Hz in Europe). Natural sources include transients from lightning, but also permanent fluctuations from Schumann resonances, which are pumped by electric discharges all over the world \cite{KoEA2017}. The EM fluctuations do not necessarily need to occur in the GW detection band since they can also interfere with detector control relying on signals at MHz, or non-linear couplings can produce up- and down-converted noise. Some of the EM noise can also depend on the environment, e.g., especially underground it is possible that magnetic properties of the surrounding rock lead to (de)amplification of natural field fluctuations \cite{AtEA2016}, which can also change with moisture content and temperature of the rock.

It is clear that due to the large variety of sources, fluctuations should be expected to vary significantly over all time scales from very brief, strong transients, to yearly seasonal cycles of, for example, Schumann resonances and local changes in rock properties. As we will show, if field fluctuations in the environment (natural or produced by the electronic infrastructure) of ET were as they are today at existing detectors, and if these fluctuations coupled as strongly with the detector output as they do in existing detectors, then ET's main environmental noise would likely be of electromagnetic origin. 

Two strategies can in principle greatly reduce problems arising from EM disturbances: (1) electronics are designed to minimize EM coupling between its components and with the environment as much as possible, (2) electronics are designed to produce the weakest possible EM disturbances. If this is achieved successfully, probably as a result of a long-lasting detector commissioning process, then the remaining problem is the unavoidable coupling to natural fluctuations, for example, because of magnetic components of the actuation system. Among all sources, the Schumann resonances play an important role since they can lead to correlated noise in a global detector network \cite{TCS2013}. It was proposed to apply noise-cancellation techniques to reduce noise from Schumann resonances \cite{CoEA2018c}.

For the model shown in Figure \ref{fig:envnoise}, we used a fit to the natural background of magnetic fluctuations associated with Schumann resonances \cite{CoEA2016b},
\begin{equation}
B=6\cdot10^{-14} / \sqrt{f/10\,{\rm Hz}}\,\rm T/\sqrt{Hz},
\end{equation}
which is about two orders of magnitude weaker than the actually measured magnetic fluctuations inside the Virgo buildings \cite{CiEA2018}. The coupling of these fluctuations with the detector output is taken from Virgo measurements \cite{CiEA2019} (similar coupling obtained at LIGO \cite{AbEA2016f}),
\begin{equation}
c = 3.3\cdot 10^{-8}/\left(f/10\,{\rm Hz}\right)^{2.8}\,\rm m/T,
\end{equation}
lowered by the (foreseen) ratio of test masses between Virgo and ET, 42/211 \cite{AcEA2015,HiEA2011}, which assumes that magnetic noise enters as test-mass displacement noise. Other coupling mechanisms, less well understood, might be important. It is also assumed that magnetic noise from Schumann resonances does not experience significant common-mode rejection due to potential differences in the coupling strength at different test masses. We use the same spectrum of magnetic fluctuations and the same coupling at all test masses.

\subsection{Environmental noise as site-selection criterion}
\begin{figure}[ht!]
\includegraphics[width=0.9\columnwidth]{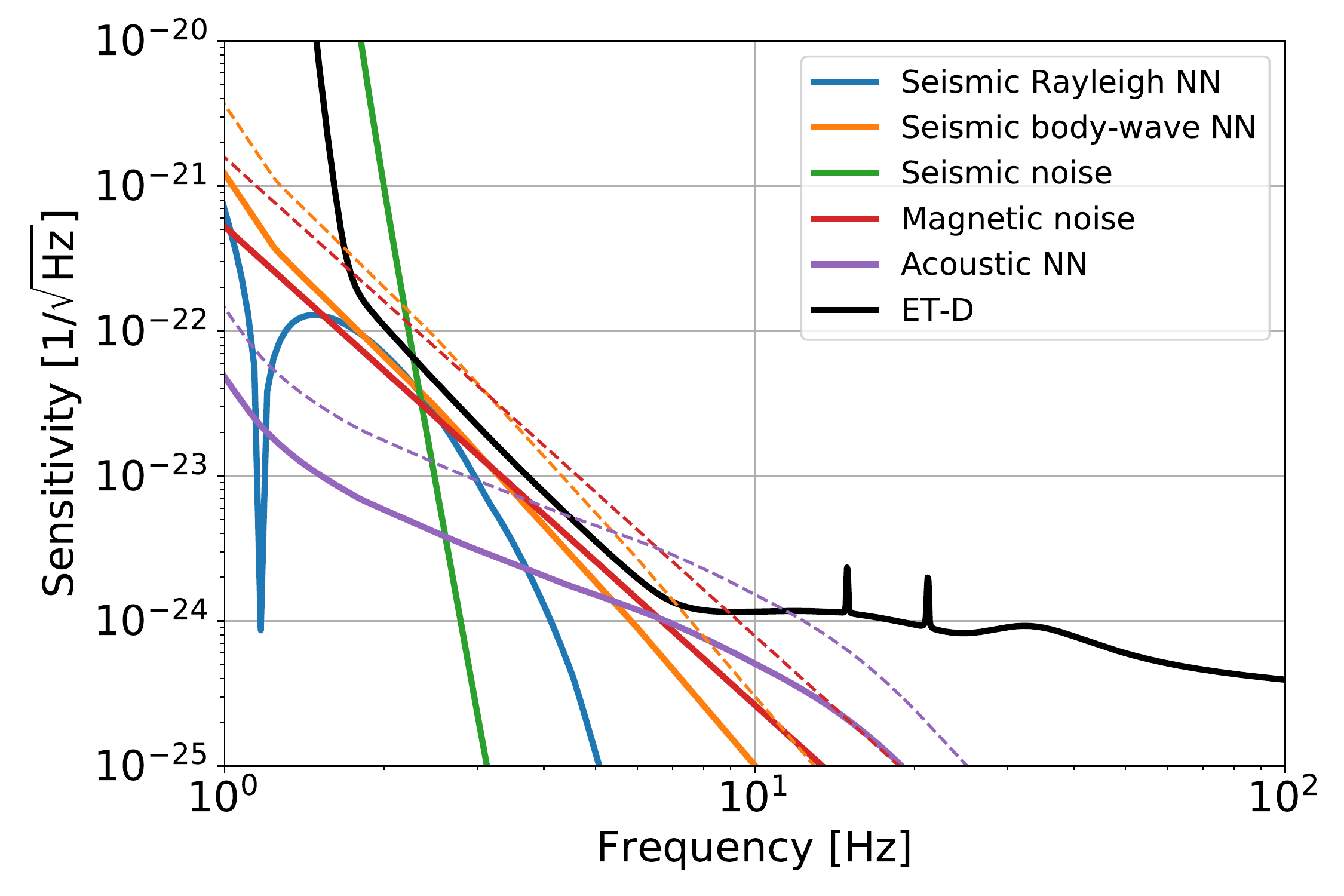}
\caption{Example of an ET environmental-noise budget together with the latest ET sensitivity model \cite{HiEA2011}. Dashed lines indicate noise levels without the required additional noise mitigation (factor 3 in all three cases), for example, by noise cancellation. It is assumed that the detector depth is 300\,m.}
\label{fig:envnoise}
\end{figure}

It is difficult to anticipate the full impact environmental noise will have on ET. A feasible task, which is also most important to ET's science potential, is to evaluate the direct environmental impact on detector sensitivity. Doing this for the two candidate sites with the equations provided in this paper, one values the site more highly that produces less environmental noise. A summary of selection parameters is shown in figure \ref{fig:envnoise} using couplings and noise models given in the previous sections. However, the detector commissioning needs to address a much wider class of coupling mechanisms and environmental influences typically involving detector control, but also, for example, up-conversion of low-frequency seismic motion in scattered-light noise. These forms of environmental noise depend strongly on the mechanical and optical engineering, like the implementation of baffles to block stray light or reduction of readout noise of optical sensors used for control, which is why we have not attempted to include these contributions in our noise budget. They will certainly have to be addressed in the technical design of ET. Generally, there is the expectation that modern control and environmental monitoring techniques involving machine learning and robotics might eventually play an important role in providing enhanced immunity of a detector to environmental influences \cite{HMS2002,CSG2018,MuEA2019,CoEA2019}. 

The approach here is to consider the simplified problem of direct environmental coupling, and therefore to use a noise budget as in figure \ref{fig:envnoise} to evaluate a site. In this sense, it is favorable to choose a site with lowest levels of environmental disturbances (low seismic and acoustic noise, weak wind, ...), but other factors may be important.

Concerning the underground NN estimates, it is favorable to have strong suppression with depth. In the case of seismic NN, this would be the case if the speed of Rayleigh waves is low. However, there is a trade-off since low-speed sites also typically show higher levels of seismic noise \cite{AkRi2009}, because a seismic source exerting a force onto the ground creates displacement amplitudes propotional to $1/(\rho c^2)$, where $\rho$ is the density of the ground, and $c$ stands for the compressional or shear-wave speeds. The two effects compensate to some extent. At sites with homogeneous geology, stiffer rock leads to an overall advantage in terms of underground NN, but soil layering can provide additional NN reduction underground so that it is not immediately clear if ultimately a typical low-speed or high-speed site is favorable. The best way to decide is by directly comparing NN estimates; dispersion curves and seismic spectra are its two most important ingredients. We note that seismic speed has no significant impact on body-wave NN.

It can be argued that some short-comings of a site in terms of seismic NN can be compensated by NN cancellation. This is certainly true, but unlike for surface detectors, cancellation of NN from a body-wave field is much more challenging as shown in \cite{BaHa2019}, where a factor 2 -- 3 of robust noise reduction was demonstrated in a simulation with 15 seismometers per test mass in a plane, isotropic, body-wave field. Based on these results, it is realistic to assume that for a factor 3 NN reduction in ET, a few tens of seismometers would be required per test mass (ET has 12 test masses in total significantly affected by NN) deployed in boreholes some of which being a few 100\,m deeper than the detector. Such a system would be a larger and more costly effort. The most challenging part would be to determine where to drill the boreholes and where to place the seismometers to achieve an effective NN reduction.

Concerning atmospheric, acoustic NN, there is currently no known technology to reduce it by noise cancellation as discussed in section \ref{sec:atmospheric}, therefore one must avoid that it contributes significantly to the ET detector noise. It is also unlikely that the acoustic field at candidate sites will be known in sufficient detail to make precise estimates of how deep the detector needs to be. Therefore, a safety margin needs to be calculated for the detector depth based on properties of the acoustic field at each candidate site to avoid any potential issue with atmospheric, acoustic NN. At depths of $\sim$300\,m, the properties of the atmosphere would not contribute to the site-selection criteria anymore. 

It is important to stress again that ET is to be understood as an infrastructure that will host a variety of detector configurations throughout its projected lifetime. Reaching the environmental noise as shown in figure \ref{fig:envnoise} will likely be a process taking many years and potentially requires major detector upgrades, but at the same time, one should not consider the predicted environmental noise as ultimate infrastructural limitation. For all noises, there may be ways of mitigation beyond the spectra shown here, but it is not possible to produce reliable predictions when the required technologies may become available.

\section{Site characterization and measurements}
\label{sec:sitechar}

\subsection{Seismic field}
A series of surface and downhole geophysical measurements need to be performed for accurate seismic noise characterization of the site in addition to providing information for geological prediction. Since seismic noise plays such a central role to environmental noise modeling, and since it has a large impact on detector infrastructure, some measurement targets must be met, while others are less important. We therefore divide the targets into "necessary" and "useful". The main targets of seismic measurements are (1) to analyze the wavefield in terms of wave propagation (dispersion, direction, amplitude) and where possible to identify local seismic sources, (2) to estimate the composition of the seismic field in terms of body waves and surface waves, (3) to assess the temporal variability of seismic sources or the seismic field. Most of this information is essential input to the noise models presented in section \ref{sec:seismic}. In addition, source identification will help to determine the size of the source-exclusion area needed around ET vertices.

\subsubsection{Necessary measurements}
\paragraph{Long-duration measurements} These measurements are aimed at characterizing the seasonal variability of the seismic ground motion spectrum \cite{HaEA2010,NaEA2014,BaEA2017,MaEA2018,VaEA2019}. Apart from variations in amplitude and peak-frequency of the oceanic microseism ($0.07-0.5$\,Hz), the temporal variation of anthropogenic noise is of utmost importance since it lies within ET's detection band. Seismic ground motion measurements on the surface and underground need to be carried out with high-class broadband, tri-axial seismometers. Downhole measurements must be carried out at depths representative of the future detector depth. The underground measurements must also be synchronized in time between themselves and with the surface measurements to obtain the cross-correlation between the two observations. Three-component measurements are also needed for computing the spectral ratio of the horizontal to vertical ground motion (H/V) at the site \cite{nakamura1989method}. The H/V ratio at the site can be used to infer source mechanisms of the noise at the site as well as information about shallow geology, for example, the basement-resonance frequency at site and the depth to bedrock \cite{acerra2004guidelines}.
        
\paragraph{Short-duration measurements} These measurements can assess more detailed spatial variations of the seismic field, as well as provide a more complex characterization of the seismic field requiring seismic arrays, for example, to infer about the body to surface wave content of the seismic noise and the propagation characteristics like the surface-wave dispersion and its propagation direction. Hence, as a second endeavor, seismic-array measurements need to be carried out in areas surrounding the detector vertices. 

Seismometers are to be chosen according to the ambient seismic-noise spectrum and should achieve a signal-to-noise ratio better than $10$ between $3$ and $10$\,Hz \cite{bensen2007processing}. If sensitivities are lower, analysis results can be strongly biased by the array's inability to provide data for correct parameter estimation of waves from short-lived seismic sources. Signal-to-noise ratios of $10$ and higher can always be achieved in surface measurements, but it might be impossible at some frequencies for underground array measurements at very quiet sites. In such cases, the SNR threshold can be reduced to $7$ profiting from the higher level of stationarity of the seismic field \cite{yang2007ambient}.

The minimum and the maximum array aperture would be based on \emph{a priori} estimates of Rayleigh-wave speeds in the same frequency band. Asten \& Henstridge, 1984 \cite{asten1984array} proposed that within a given frequency band for stochastic analysis, the maximum sensor separation $d_{\rm max}$ should be at least greater than the maximum wavelength of interest $\lambda_{\rm max}$ and the minimum sensor separation $d_{\rm min}$ must be less than half the minimum wavelength $\lambda_{\rm min}$. The second condition follows from the \textit{Nyquist} criterion to avoid spatial aliasing at smaller wavelengths. Following the above two conditions for designing surface-seismic arrays, we propose surface seismometers to be installed approximately along rings of increasing radii and equally spaced in azimuth in each ring. Studies by Kimman et al., 2012 \cite{kimman2012characteristics} and Koley et al., 2018 \cite{koley2018seismic}, which use the concept of theoretical array response \cite{woods1973plane}, have shown useful applications of such array geometries for ambient noise studies. The main target of the array measurements would be estimation of the surface wave dispersion curve, characterization of seismic sources, unravelling the anisotropy of the seismic field, and estimation of the modal content of the seismic noise. A minimal measurement period of several weeks is recommended for understanding the diurnal and the weekly variation in the seismic noise properties.  

\subsubsection{Other interesting measurements}
\begin{itemize}
\setlength\itemsep{0em}
    \item Underground measurements at all three foreseen vertex locations using high-class broadband sensors. These measurements should at least be carried out for a few weeks. The main purpose is to characterize spatial variations of the seismic field underground.
    
    \item Three-dimensional array measurements around tentative locations of detector vertices between 3\,Hz and 10\,Hz. Highest quality seismometers preferably with self-noise below Peterson's global low-noise model in the relevant frequency band are to be used at least for the underground seismometers. Some analysis results would greatly improve by using three-axis seismometers. The data can be used to provide an accurate prediction of seismic Newtonian noise using detailed information about the seismic field in terms of polarizations, propagation directions and seismic speeds of all wave types, scattering from the surface, etc. Since such array measurements are very costly, they should be planned with seismologists to maximize the science output and be carried out for a year or longer. It is opportune to make use of existing underground infrastructure \cite{MaEA2018}.
\end{itemize}

\subsubsection{Seismic Methods}

\paragraph{Passive seismic} Under a deterministic approach, the ambient seismic wavefield may be treated as a combination of plane waves, whose apparent velocity and direction of propagation may be conveniently retrieved using array processing schemes such as the frequency-wavenumber power spectral analysis \cite{KrVi1996,park1999multichannel}. This method can be applied to: (i) human noise frequency band (1--10\,Hz), which allows penetration depths on the order of 20--500\,m ; (ii) microseismic noise frequency band (0.1--1\,Hz), whose corresponding penetration depths are on the order of 500--10000\,m. The main advantage of studying the ambient field is that costly active sources are not needed. However, the method requires long-duration recordings in order to explore the full spatial distribution of noise sources. Another challenge coming with analyses of the ambient field is the separation of wave polarizations, which is important, for example, for certain techniques to determine velocity profiles. 

The properties of the seismic noise over the 1--10\,Hz frequency band at the vertices are conveniently retrieved using an array of seismometers. Array analysis allows to (i) derive the kinematic properties (i.e., direction-of-arrival, apparent velocity) of the noise wavefield, so to get inferences on the location of the main source(s), and (ii) to get information on the surface-wave dispersion function, to be finally inverted for a shallow 1D model of the shear-wave velocity at the site. By applying the $\lambda$/4 rule, these signals are correctly sampled by arrays whose apertures (largest inter-station distance) are about 80\,m. Sampling different frequency ranges would require different apertures.

For a target wavelength, in principle only three seismometers are sufficient for retrieving the kinematic properties of the incoming wavefield. Nonetheless, the higher the number of seismometers, the better will be the precision in the estimate of those parameters. In addition, if a large number of instruments is available, one may attempt to deploy an array whose density and aperture are appropriate for the entire wavelength range of interest.  The high cost of high-sensitivity  seismometers  poses however  limitations on the number of  instruments to be employed. Thus a reasonable compromise could be the installation of a 8--10 element array. The installation can be replicated at the three different vertices, or the same array moved in between the three vertices allowing 10--15 days of recording at each site. An exact determination about the time duration of recording may be provided only after a characterisation on the location and temporal variability of the main noise sources.
    
\paragraph{Active seismic} A survey based on reflection/refraction seismology can provide seismic-wave velocity profiles or geometrical information about subsurface structures \cite{Mon2010}. In its simplest form, the active survey is done deploying geophones evenly spaced along a line on the surface, and the seismic source can be a vibroseis truck or an excavator. Often, explosives are deployed in shallow boreholes. These sources produce body and surface waves, which can be studied individually. As a rule of thumb, a velocity profile can be obtained to a depth corresponding to a quarter of the length of the line connecting the seismometers. The optimal spacing between seismometers depends on the targeted spatial resolution, which should be similar to the length of the shortest waves in the frequency band of interest, i.e., higher spatial resolution is required to characterize near-surface soil determining the propagation of slow Rayleigh waves, and relatively low resolution is acceptable to characterize deeper rock, where fast body waves dominate.

\subsection{Atmospheric fields}
The importance of characterizing atmospheric fields for site selection depends on the depth of the future detector. Avoiding atmospheric NN is one of the main motivations to construct ET underground. Already at 100\,m depth, atmospheric acoustic NN is likely insignificant \cite{FiEA2009}, but as explained in Section \ref{sec:envnoise}, suppression of acoustic NN with depth strongly depends on the anisotropy of the acoustic field, and more detailed numerical studies are required to determine the minimum depth, at which acoustic NN can be safely neglected. Suppression of acoustic NN with depth also depends on two-point spatial correlations, which are influenced by source distributions and scattering of acoustic waves. Therefore, when the considered detector depth is such that a significant contribution from atmospheric, acoustic NN cannot be ruled out, sound spectra, propagation directions and spatial correlations measured with microphone arrays are important site-characterization targets. These should be deployed at the surface of all foreseen vertex locations, and the required number of microphones for the analysis of the ambient acoustic field is the same as for the seismic measurements, i.e., several tens of sensors are recommended, but a handful of sensors is already sufficient to carry out velocity measurements and to determine propagation directions.

Good quality acoustic measurements are challenging in open environments due to wind noise. The usage of wind shields, and averaging microphone signals over some number of nearby microphones are straight-forward strategies to lower wind noise \cite{WaHe2009,NoEA2014}. The impact of wind noise on sound spectra can always be assessed by calculating cross-spectral densities between two nearby microphones.

Another measurement target is average wind speed since it is the main parameter influencing the suppression of advection NN with depth. It is also important to consider the surface structure and whether wind might lead to vortices of the right scale that could lower the suppression of advection NN with depth. The best way to estimate advection NN at a site is to deploy a LIDAR system. It can be used to make volumetric measurements of temperature, humidity, and wind fields \cite{CLN2004,Beh2005,HaEA2015b,SpEA2016}, but different LIDAR systems are sensitive to different variables, which means that several LIDAR systems may be used. Again, deployment of such a system should be at the foreseen vertex locations, and to carry out velocity, temperature and humidity measurements for as long as possible (ideally a year), but even brief measurements would provide a wealth of data useful to improve advection NN models.

\subsection{Electromagnetic field}
As we have seen in Section \ref{sec:envnoise}, the electromagnetic field, especially fluctuations of the magnetic field, play a very important role in ET, and they require attention. It is however difficult to assess this form of environmental noise in advance since the EM field will likely be dominated by sources installed with the detector and its infrastructure. The main motivation to carry out measurements of the (electro)magnetic field as part of a site-selection campaign is to make sure that there is not an abundance of EM transients from local sources like nearby power lines or transformer stations. These measurements should be carried out at all three foreseen vertex locations.

The Schumann resonances have similar spectra everywhere on Earth, which means that they are a minor item of site evaluation. A characterization of local, natural sources such as lightning strikes can be done, but is not likely to significantly contribute to the science criteria for site selection. If underground measurements are possible, then a comparison of surface and underground Schumann resonances can reveal local magnetic amplifications by the surrounding rock. For the observation of Schumann resonances, high-quality, induction-coil magnetometers are required, ideally buried to avoid noise from wind-induced vibrations.

\subsection{Geotechnical, geographic, and other surveys}
Geotechnical investigations are key to any tunnel construction, typically contributing 2\% -- 7\% to the total construction cost \cite{Hem2012}. It is largely based on analyses of the surface, e.g., outcrops, and of drill cores at the construction site. For deep sites, it has to be accompanied by geophysical studies, for example, to investigate sub-surface geology and reduce the uncertainty of the geological models. Exploratory boring averages about 1.5\,m of borehole per tunnel meter \cite{Hem2012}. Detailed information of (hydro)geological and groundwater conditions are essential to plan the construction and estimate the construction cost, and to foresee potential issues with the presence of water and water handling during detector operation. Possible values of rock permeability to water span ten orders of magnitude, which makes groundwater conditions especially difficult to predict \cite{Par2004}. Conditions can also change significantly with season. Incompleteness of information can lead to delays in construction and increased cost, sometimes even to major construction failure \cite{LiEA2015}. A thorough geotechnical survey is necessary for a smooth construction process, but it never provides a guarantee against unforeseen problems since geological conditions can change over small distances. A historical collection of tunnel construction cost can be found in Rostami et al \cite{RoEA2013}.

However, since these investigations are very costly, they cannot be carried out in their full extent at both ET candidate sites. Instead, in preparation of a site selection, only a small number of boreholes can be realized to provide enough information for a site selection, not for a detailed cost estimate and construction planning. The information provided by these preliminary geotechnical investigations include stratigraphy, elevation of the groundwater table, limited information on rock quality, and some idea of how these parameters vary over the area of interest. This information can help to refine models of environmental noise, but also provide important input for approximate construction cost estimates.

Another set of site studies concerns the collection of already available data or potentially easily retrievable data about weather, geomorphology, a geodatabase, orthophotos, digital elevation models, land use, parks and protected areas, hazard maps, and hydrology of the region. Data can also be available about crustal deformation and ground stability, e.g., subsidence and shear, from past DInSAR analyses \cite{ToEA2014}, or installations of GNSS stations \cite{JRH2017}. Some understanding of ground stability is of course crucial for site selection. Additional hydrological data can be obtained by groundwater well extraction, piezometers, and pumping tests.

\section{Conclusion}
\label{sec:conclusion}
This paper provides an overall assessment of site-selection criteria for the proposed next-generation, underground GW detector ET, and gives guidelines for site-characterization campaigns and noise modeling. Its main purpose is to inform the ET and broader science communities about the main challenges in the preparation of a site selection. It is important to understand how strongly the quality of the ET infrastructure in terms of lifetime and science potential depends on site conditions. Early understanding of the short-comings of a site can help to devise technological solutions to overcome certain limitations.

The very large number of individual site parameters demonstrates the complexity of a site evaluation. Detector lifetime, operation and sensitivity are of prime interest to the project, but construction and operation cost might be the decisive factors for site selection. Given the scale of the investment, it is also clear that the socio-economic impact of ET will be considered and will play an important role.

As for many other modern experiments in fundamental physics, the environment can have a significant impact on the science potential of the ET research infrastructure. In fact, the main reason to construct ET underground, and therefore the main contribution to construction cost, is to avoid environmental noise from terrestrial gravity fluctuations associated with atmospheric and surface seismic fields. However, even underground the observation band of ET can be limited by environmental noise, which means that noise modeling forms an essential part of the site evaluation. We presented a formalism to project observations of environmental noise, such as seismic displacement and acoustic noise, into ET instrument noise, and we conclude that all forms of ambient noise can potentially limit ET sensitivity.

The advantage of having a high-quality, low-noise site means that more care needs to be taken to preserve site quality over the envisioned $\gtrsim 50$ years of ET lifetime. This can be achieved by negotiating kilometer-scale protective areas around the three vertex locations of ET preventing, for example, new industry, roads or railways to introduce disturbances. 

All these considerations are key to the planning of a site-characterization campaign and to obtain a site evaluation. In the end, the value of a site will not only depend on its properties, but also on the proposed solutions to address challenges specific to a site. 

\begin{acknowledgments}
Part of this research was conducted by the Australian Research Council Centre of Excellence for Gravitational Wave Discovery (OzGrav), through project number CE170100004. The work in Hungary was supported by the grant National Research, Development and Innovation Office –  NKFIH 124366(124508). The support of the European Regional Development Fund and of Hungary in the frame of the project GINOP-2.2.1-15-2016-00012 is acknowledged. Part of the Italian contribution is funded by INFN thanks to the “Protocollo di Intesa il Ministero dell’Istruzione, dell’Università e della Ricerca, la Regione Autonoma della Sardegna, l’Istituto Nazionale di Fisica Nucleare e l’Università degli Studi di Sassari finalizzato a sostenere la candidatura italiana a ospitare l’infrastruttura Einstein Telescope in Sardegna e al potenziamento di VIRGO” (2018). The contribution of the University of Sassari is funded by the FSC 2014-2020 – Patto per lo Sviluppo della Regione Sardegna. The Spanish contribution is funded by the State Research Agency, Ministry of Science, Innovation and Universities (grant n. FPA2016-76821-P), European Union FEDER funds, and Vicepresid\`encia i Conselleria d'Innovaci\'o, Recerca i Turisme del Govern de les Illes Balears. TB was supported by the TEAM/2016-3/19 grant from FNP.

The data that support the findings of this study are available from the corresponding author upon reasonable request.
\end{acknowledgments}

\bibliographystyle{aipsamp}
\bibliography{references}

\providecommand{\newblock}{}
\begin{thebibliography}{100}
\expandafter\ifx\csname url\endcsname\relax
  \def\url#1{{\tt #1}}\fi
\expandafter\ifx\csname urlprefix\endcsname\relax\def\urlprefix{URL }\fi
\providecommand{\eprint}[2][]{\url{#2}}

\bibitem{ArEA2013a}
Armengaud E, Augier C, Beno{\^i}t A, Beno{\^i}t A, Bergé L, Bergmann T,
  Blümer J, Broniatowski A, Brudanin V, Censier B, Chapellier M, Charlieux F,
  Couedo F, Coulter P, Cox G, Jesus M~D, Domange J, Drilien A~A, Dumoulin L,
  Eitel K, Filosofov D, Fourches N, Gascon J, Gerbier G, Gros M, Henry S,
  Hervé S, Heuermann G, Holtzer N, Juillard A, Kleifges M, Kluck H, Kozlov V,
  Kraus H, Kudryavtsev V, Sueur H~L, Loaiza P, Marnieros S, Menshikov A, Navick
  X~F, Nones C, Olivieri E, Pari P, Paul B, Rigaut O, Robinson M, Rozov S,
  Sanglard V, Schmidt B, Scorza S, Siebenborn B, Semikh S, Tcherniakhovski D,
  Torrento-Coello A, Vagneron L, Walker R, Weber M, Yakushev E and Zhang X 2013
  {\em Astroparticle Physics\/} {\bf 47} 1 -- 9 ISSN 0927-6505
  \urlprefix\url{http://www.sciencedirect.com/science/article/pii/S0927650513000790}

\bibitem{ApEA2011}
Aprile E, Arisaka K, Arneodo F, Askin A, Baudis L, Behrens A, Bokeloh K, Brown
  E, Cardoso J~M~R, Choi B, Cline D, Fattori S, Ferella A~D, Giboni K~L, Kish
  A, Lam C~W, Lamblin J, Lang R~F, Lim K~E, Lin Q, Lindemann S, Lindner M,
  Lopes J~A~M, Lung K, Marrod\'an~Undagoitia T, Mei Y, Melgarejo~Fernandez A~J,
  Ni K, Oberlack U, Orrigo S~E~A, Pantic E, Plante G, Ribeiro A~C~C, Santorelli
  R, dos Santos J~M~F, Schumann M, Shagin P, Simgen H, Teymourian A, Thers D,
  Tziaferi E, Wang H, Weber M and Weinheimer C (XENON100 Collaboration) 2011
  {\em Phys. Rev. D\/} {\bf 83}(8) 082001
  \urlprefix\url{https://link.aps.org/doi/10.1103/PhysRevD.83.082001}

\bibitem{AgEA2014}
Agostini M, Allardt M, Andreotti E, Bakalyarov A~M, Balata M, Barabanov I,
  Barnab{\'e}~Heider M, Barros N, Baudis L, Bauer C, Becerici-Schmidt N,
  Bellotti E, Belogurov S, Belyaev S~T, Benato G, Bettini A, Bezrukov L, Bode
  T, Brudanin V, Brugnera R, Budj{\'a}{\v{s}} D, Caldwell A, Cattadori C,
  Chernogorov A, Cossavella F, Demidova E~V, Domula A, Egorov V, Falkenstein R,
  Ferella A, Freund K, Frodyma N, Gangapshev A, Garfagnini A, Gotti C, Grabmayr
  P, Gurentsov V, Gusev K, Guthikonda K~K, Hampel W, Hegai A, Heisel M, Hemmer
  S, Heusser G, Hofmann W, Hult M, Inzhechik L~V, Ioannucci L, Cs{\'a}thy J
  Janicsk{\'o}and~Jochum J, Junker M, Kihm T, Kirpichnikov I~V, Kirsch A,
  Klimenko A, Kn{\"o}pfle K~T, Kochetov O, Kornoukhov V~N, Kuzminov V~V,
  Laubenstein M, Lazzaro A, Lebedev V~I, Lehnert B, Liao H~Y, Lindner M, Lippi
  I, Liu X, Lubashevskiy A, Lubsandorzhiev B, Lutter G, Macolino C, Machado
  A~A, Majorovits B, Maneschg W, Nemchenok I, Nisi S, O'Shaughnessy C,
  Palioselitis D, Pandola L, Pelczar K, Pessina G, Pullia A, Riboldi S, Sada C,
  Salathe M, Schmitt C, Schreiner J, Schulz O, Schwingenheuer B, Sch{\"o}nert
  S, Shevchik E, Shirchenko M, Simgen H, Smolnikov A, Stanco L, Strecker H,
  Tarka M, Ur C~A, Vasenko A~A, Volynets O, von Sturm K, Wagner V, Walter M,
  Wegmann A, Wester T, Wojcik M, Yanovich E, Zavarise P, Zhitnikov I, Zhukov
  S~V, Zinatulina D, Zuber K and Zuzel G 2014 {\em The European Physical
  Journal C\/} {\bf 74} 2764 ISSN 1434-6052
  \urlprefix\url{https://doi.org/10.1140/epjc/s10052-014-2764-z}

\bibitem{AkEA2015}
Akerib D, Araújo H, Bai X, Bailey A, Balajthy J, Bernard E, Bernstein A,
  Bradley A, Byram D, Cahn S, Carmona-Benitez M, Chan C, Chapman J, Chiller A,
  Chiller C, Coffey T, Currie A, de~Viveiros L, Dobi A, Dobson J, Druszkiewicz
  E, Edwards B, Faham C, Fiorucci S, Flores C, Gaitskell R, Gehman V, Ghag C,
  Gibson K, Gilchriese M, Hall C, Hertel S, Horn M, Huang D, Ihm M, Jacobsen R,
  Kazkaz K, Knoche R, Larsen N, Lee C, Lindote A, Lopes M, Malling D, Mannino
  R, McKinsey D, Mei D~M, Mock J, Moongweluwan M, Morad J, Murphy A, Nehrkorn
  C, Nelson H, Neves F, Ott R, Pangilinan M, Parker P, Pease E, Pech K, Phelps
  P, Reichhart L, Shutt T, Silva C, Solovov V, Sorensen P, O’Sullivan K,
  Sumner T, Szydagis M, Taylor D, Tennyson B, Tiedt D, Tripathi M, Uvarov S,
  Verbus J, Walsh N, Webb R, White J, Witherell M, Wolfs F, Woods M and Zhang C
  2015 {\em Astroparticle Physics\/} {\bf 62} 33 -- 46 ISSN 0927-6505
  \urlprefix\url{http://www.sciencedirect.com/science/article/pii/S0927650514001054}

\bibitem{AlEA2015}
Albert J~B, Auty D~J, Barbeau P~S, Beck D, Belov V, Benitez-Medina C,
  Breidenbach M, Brunner T, Burenkov A, Cao G~F, Chambers C, Cleveland B, Coon
  M, Craycraft A, Daniels T, Danilov M, Daugherty S~J, Davis C~G, Davis J,
  Delaquis S, Der Mesrobian-Kabakian A, DeVoe R, Didberidze T, Dolgolenko A,
  Dolinski M~J, Dunford M, Fairbank W, Farine J, Feldmeier W, Fierlinger P,
  Fudenberg D, Giroux G, Gornea R, Graham K, Gratta G, Hall C, Herrin S, Hughes
  M, Jewell M~J, Jiang X~S, Johnson A, Johnson T~N, Johnston S, Karelin A,
  Kaufman L~J, Killick R, Koffas T, Kravitz S, Kuchenkov A, Kumar K~S, Leonard
  D~S, Licciardi C, Lin Y~H, Ling J, MacLellan R, Marino M~G, Mong B, Moore D,
  Nelson R, Odian A, Ostrovskiy I, Piepke A, Pocar A, Prescott C~Y, Rivas A,
  Rowson P~C, Russell J~J, Schubert A, Sinclair D, Smith E, Stekhanov V, Tarka
  M, Tolba T, Tsang R, Twelker K, Vuilleumier J~L, Waite A, Walton J, Walton T,
  Weber M, Wen L~J, Wichoski U, Wood J, Yang L, Yen Y~R and Zeldovich O~Y
  (EXO-200 Collaboration) 2015 {\em Phys. Rev. C\/} {\bf 92}(1) 015503
  \urlprefix\url{https://link.aps.org/doi/10.1103/PhysRevC.92.015503}

\bibitem{AlEA2017}
Alduino C, Alfonso K, Artusa D~R, Avignone F~T, Azzolini O, Banks T~I, Bari G,
  Beeman J~W, Bellini F, Benato G, Bersani A, Biassoni M, Branca A, Brofferio
  C, Bucci C, Camacho A, Caminata A, Canonica L, Cao X~G, Capelli S, Cappelli
  L, Carbone L, Cardani L, Carniti P, Casali N, Cassina L, Chiesa D, Chott N,
  Clemenza M, Copello S, Cosmelli C, Cremonesi O, Creswick R~J, Cushman J~S,
  D'Addabbo A, Dafinei I, Davis C~J, Dell'Oro S, Deninno M~M, Di~Domizio S,
  Di~Vacri M~L, Drobizhev A, Fang D~Q, Faverzani M, Fernandes G, Ferri E,
  Ferroni F, Fiorini E, Franceschi M~A, Freedman S~J, Fujikawa B~K, Giachero A,
  Gironi L, Giuliani A, Gladstone L, Gorla P, Gotti C, Gutierrez T~D, Haller
  E~E, Han K, Hansen E, Heeger K~M, Hennings-Yeomans R, Hickerson K~P, Huang
  H~Z, Kadel R, Keppel G, Kolomensky Y~G, Leder A, Ligi C, Lim K~E, Ma Y~G,
  Maino M, Marini L, Martinez M, Maruyama R~H, Mei Y, Moggi N, Morganti S,
  Mosteiro P~J, Napolitano T, Nastasi M, Nones C, Norman E~B, Novati V,
  Nucciotti A, O'Donnell T, Ouellet J~L, Pagliarone C~E, Pallavicini M,
  Palmieri V, Pattavina L, Pavan M, Pessina G, Pettinacci V, Piperno G, Pira C,
  Pirro S, Pozzi S, Previtali E, Rosenfeld C, Rusconi C, Sakai M, Sangiorgio S,
  Santone D, Schmidt B, Schmidt J, Scielzo N~D, Singh V, Sisti M, Smith A~R,
  Taffarello L, Tenconi M, Terranova F, Tomei C, Trentalange S, Vignati M,
  Wagaarachchi S~L, Wang B~S, Wang H~W, Welliver B, Wilson J, Winslow L~A, Wise
  T, Woodcraft A, Zanotti L, Zhang G~Q, Zhu B~X, Zimmermann S, Zucchelli S and
  Laubenstein M 2017 {\em The European Physical Journal C\/} {\bf 77} 543

\bibitem{TOEA2010}
Thomas-Osip J~E, McCarthy P, Prieto G, Phillips M~M and Johns M 2010 {Giant
  Magellan Telescope site testing: summary} {\em Ground-based and Airborne
  Telescopes III\/} vol 7733 ed Stepp L~M, Gilmozzi R and Hall H~J
  International Society for Optics and Photonics (SPIE) pp 569 -- 578
  \urlprefix\url{https://doi.org/10.1117/12.856934}

\bibitem{RiEA2008}
Riddle R~L, Walker D, Schöck M, Els S~G, Skidmore W, Travouillon T, Bustos E,
  Seguel J, Vasquez J, Blum R~D, Gillett P and Gregory B 2008 {An analysis of
  light pollution at the Thirty Meter Telescope candidate sites} {\em
  Ground-based and Airborne Telescopes II\/} vol 7012 ed Stepp L~M and Gilmozzi
  R International Society for Optics and Photonics (SPIE) pp 850 -- 861
  \urlprefix\url{https://doi.org/10.1117/12.787295}

\bibitem{VeEA2011}
Vernin J, Mu{\~{n}}oz-Tu{\~{n}}{\'{o}}n C, Sarazin M, Rami{\'{o}} H~V, Varela
  A~M, Trinquet H, Delgado J~M, Fuensalida J~J, Reyes M, Benhida A, Benkhaldoun
  Z, Lambas D~G, Hach Y, Lazrek M, Lombardi G, Navarrete J, Recabarren P, Renzi
  V, Sabil M and Vrech R 2011 {\em Publications of the Astronomical Society of
  the Pacific\/} {\bf 123} 1334--1346
  \urlprefix\url{https://doi.org/10.1086\%2F662995}

\bibitem{VaEA2012}
Rami{\'{o}} H~V, Vernin J, Mu{\~{n}}oz-Tu{\~{n}}{\'{o}}n C, Sarazin M, Varela
  A~M, Trinquet H, Delgado J~M, Fuensalida J~J, Reyes M, Benhida A, Benkhaldoun
  Z, Lambas D~G, Hach Y, Lazrek M, Lombardi G, Navarrete J, Recabarren P, Renzi
  V, Sabil M and Vrech R 2012 {\em Publications of the Astronomical Society of
  the Pacific\/} {\bf 124} 868--884
  \urlprefix\url{https://doi.org/10.1086\%2F667599}

\bibitem{VaEA2014}
Varela A~M, Rami{\'{o}} H~V, Vernin J, Mu{\~{n}}oz-Tu{\~{n}}{\'{o}}n C, Sarazin
  M, Trinquet H, Delgado J~M, Fuensalida J~J, Reyes M, Benhida A, Benkhaldoun
  Z, Lambas D~G, Hach Y, Lazrek M, Lombardi G, Navarrete J, Recabarren P, Renzi
  V, Sabil M and Vrech R 2014 {\em Publications of the Astronomical Society of
  the Pacific\/} {\bf 126} 412--431
  \urlprefix\url{https://doi.org/10.1086\%2F676135}

\bibitem{dJEA2010}
de~Jong M 2010 {\em Nuclear Instruments and Methods in Physics Research Section
  A: Accelerators, Spectrometers, Detectors and Associated Equipment\/} {\bf
  623} 445 -- 447 ISSN 0168-9002 1st International Conference on Technology and
  Instrumentation in Particle Physics
  \urlprefix\url{http://www.sciencedirect.com/science/article/pii/S0168900210005954}

\bibitem{AaEA2017}
Aartsen M, Ackermann M, Adams J, Aguilar J, Ahlers M, Ahrens M, Altmann D,
  Andeen K, Anderson T, Ansseau I, Anton G, Archinger M, Argüelles C, Auer R,
  Auffenberg J, Axani S, Baccus J, Bai X, Barnet S, Barwick S, Baum V, Bay R,
  Beattie K, Beatty J, Tjus J~B, Becker K~H, Bendfelt T, BenZvi S, Berley D,
  Bernardini E, Bernhard A, Besson D, Binder G, Bindig D, Bissok M, Blaufuss E,
  Blot S, Boersma D, Bohm C, Börner M, Bos F, Bose D, Böser S, Botner O,
  Bouchta A, Braun J, Brayeur L, Bretz H~P, Bron S, Burgman A, Burreson C,
  Carver T, Casier M, Cheung E, Chirkin D, Christov A, Clark K, Classen L,
  Coenders S, Collin G, Conrad J, Cowen D, Cross R, Day C, Day M,
  de~Andr{\'{e}} J, Clercq C~D, del Pino~Rosendo E, Dembinski H, Ridder S~D,
  Descamps F, Desiati P, de~Vries K, de~Wasseige G, de~With M, DeYoung T,
  D{\'{\i}}az-V{\'{e}}lez J, di~Lorenzo V, Dujmovic H, Dumm J, Dunkman M,
  Eberhardt B, Edwards W, Ehrhardt T, Eichmann B, Eller P, Euler S, Evenson P,
  Fahey S, Fazely A, Feintzeig J, Felde J, Filimonov K, Finley C, Flis S,
  Fösig C~C, Franckowiak A, Fr{\`{e}}re M, Friedman E, Fuchs T, Gaisser T,
  Gallagher J, Gerhardt L, Ghorbani K, Giang W, Gladstone L, Glauch T, Glowacki
  D, Glüsenkamp T, Goldschmidt A, Gonzalez J, Grant D, Griffith Z, Gustafsson
  L, Haack C, Hallgren A, Halzen F, Hansen E, Hansmann T, Hanson K, Haugen J,
  Hebecker D, Heereman D, Helbing K, Hellauer R, Heller R, Hickford S, Hignight
  J, Hill G, Hoffman K, Hoffmann R, Hoshina K, Huang F, Huber M, Hulth P,
  Hultqvist K, In S, Inaba M, Ishihara A, Jacobi E, Jacobsen J, Japaridze G,
  Jeong M, Jero K, Jones A, Jones B, Joseph J, Kang W, Kappes A, Karg T, Karle
  A, Katz U, Kauer M, Keivani A, Kelley J, Kemp J, Kheirandish A, Kim J, Kim M,
  Kintscher T, Kiryluk J, Kitamura N, Kittler T, Klein S, Kleinfelder S, Kleist
  M, Kohnen G, Koirala R, Kolanoski H, Konietz R, Köpke L, Kopper C, Kopper S,
  Koskinen D, Kowalski M, Krasberg M, Krings K, Kroll M, Krückl G, Krüger C,
  Kunnen J, Kunwar S, Kurahashi N, Kuwabara T, Labare M, Laihem K, Landsman H,
  Lanfranchi J, Larson M, Lauber F, Laundrie A, Lennarz D, Leich H,
  Lesiak-Bzdak M, Leuermann M, Lu L, Ludwig J, Lünemann J, Mackenzie C, Madsen
  J, Maggi G, Mahn K, Mancina S, Mandelartz M, Maruyama R, Mase K, Matis H,
  Maunu R, McNally F, McParland C, Meade P, Meagher K, Medici M, Meier M, Meli
  A, Menne T, Merino G, Meures T, Miarecki S, Minor R, Montaruli T, Moulai M,
  Murray T, Nahnhauer R, Naumann U, Neer G, Newcomb M, Niederhausen H, Nowicki
  S, Nygren D, Pollmann A~O, Olivas A, O{\textquotesingle}Murchadha A,
  Palczewski T, Pandya H, Pankova D, Patton S, Peiffer P, Penek O, Pepper J,
  de~los Heros C~P, Pettersen C, Pieloth D, Pinat E, Price P, Przybylski G,
  Quinnan M, Raab C, Rädel L, Rameez M, Rawlins K, Reimann R, Relethford B,
  Relich M, Resconi E, Rhode W, Richman M, Riedel B, Robertson S, Rongen M,
  Roucelle C, Rott C, Ruhe T, Ryckbosch D, Rysewyk D, Sabbatini L, Herrera S~S,
  Sandrock A, Sandroos J, Sandstrom P, Sarkar S, Satalecka K, Schlunder P,
  Schmidt T, Schoenen S, Schöneberg S, Schukraft A, Schumacher L, Seckel D,
  Seunarine S, Solarz M, Soldin D, Song M, Spiczak G, Spiering C, Stanev T,
  Stasik A, Stettner J, Steuer A, Stezelberger T, Stokstad R, Stö{\ss}l A,
  Ström R, Strotjohann N, Sulanke K~H, Sullivan G, Sutherland M, Taavola H,
  Taboada I, Tatar J, Tenholt F, Ter-Antonyan S, Terliuk A, Te{\v{s}}i{\'{c}}
  G, Thollander L, Tilav S, Toale P, Tobin M, Toscano S, Tosi D, Tselengidou M,
  Turcati A, Unger E, Usner M, Vandenbroucke J, van Eijndhoven N, Vanheule S,
  van Rossem M, van Santen J, Vehring M, Voge M, Vogel E, Vraeghe M, Wahl D,
  Walck C, Wallace A, Wallraff M, Wandkowsky N, Weaver C, Weiss M, Wendt C,
  Westerhoff S, Wharton D, Whelan B, Wickmann S, Wiebe K, Wiebusch C, Wille L,
  Williams D, Wills L, Wisniewski P, Wolf M, Wood T, Woolsey E, Woschnagg K, Xu
  D, Xu X, Xu Y, Yanez J, Yodh G, Yoshida S and Zoll M 2017 {\em Journal of
  Instrumentation\/} {\bf 12} P03012--P03012
  \urlprefix\url{https://doi.org/10.1088\%2F1748-0221\%2F12\%2F03\%2Fp03012}

\bibitem{dMu2019}
Di~Murro V 2019 {Long-term performance of a concrete-lined tunnel at CERN}
  \urlprefix\url{http://cds.cern.ch/record/2685151}

\bibitem{LIGO1989}
Vogt R, Raab F, Drever R, Thorne K and Weiss R 1989 {\em LIGO Document
  Server\/} {\bf M890001} 1--352
  \urlprefix\url{https://dcc.ligo.org/M890001/public}

\bibitem{AcEA2012}
Accadia T, Acernese F, Alshourbagy M, Amico P, Antonucci F, Aoudia S, Arnaud N,
  Arnault C, Arun K~G, Astone P, Avino S, Babusci D, Ballardin G, Barone F,
  Barrand G, Barsotti L, Barsuglia M, Basti A, Bauer T~S, Beauville F, Bebronne
  M, Bejger M, Beker M~G, Bellachia F, Belletoile A, Beney J~L, Bernardini M,
  Bigotta S, Bilhaut R, Birindelli S, Bitossi M, Bizouard M~A, Blom M, Boccara
  C, Boget D, Bondu F, Bonelli L, Bonnand R, Boschi V, Bosi L, Bouedo T, Bouhou
  B, Bozzi A, Bracci L, Braccini S, Bradaschia C, Branchesi M, Briant T,
  Brillet A, Brisson V, Brocco L, Bulik T, Bulten H~J, Buskulic D, Buy C,
  Cagnoli G, Calamai G, Calloni E, Campagna E, Canuel B, Carbognani F, Carbone
  L, Cavalier F, Cavalieri R, Cecchi R, Cella G, Cesarini E, Chassande-Mottin
  E, Chatterji S, Chiche R, Chincarini A, Chiummo A, Christensen N, Clapson
  A~C, Cleva F, Coccia E, Cohadon P~F, Colacino C~N, Colas J, Colla A,
  Colombini M, Conforto G, Corsi A, Cortese S, Cottone F, Coulon J~P, Cuoco E,
  D{\textquotesingle}Antonio S, Daguin G, Dari A, Dattilo V, David P~Y, Davier
  M, Day R, Debreczeni G, Carolis G~D, Dehamme M, Fabbro R~D, Pozzo W~D, del
  Prete M, Derome L, Rosa R~D, DeSalvo R, Dialinas M, Fiore L~D, Lieto A~D,
  Emilio M~D~P, Virgilio A~D, Dietz A, Doets M, Dominici P, Dominjon A, Drago
  M, Drezen C, Dujardin B, Dulach B, Eder C, Eleuteri A, Enard D, Evans M,
  Fabbroni L, Fafone V, Fang H, Ferrante I, Fidecaro F, Fiori I, Flaminio R,
  Forest D, Forte L~A, Fournier J~D, Fournier L, Franc J, Francois O, Frasca S,
  Frasconi F, Freise A, Gaddi A, Galimberti M, Gammaitoni L, Ganau P, Garnier
  C, Garufi F, G{\'{a}}sp{\'{a}}r M~E, Gemme G, Genin E, Gennai A, Gennaro G,
  Giacobone L, Giazotto A, Giordano G, Giordano L, Girard C, Gouaty R, Grado A,
  Granata M, Granata V, Grave X, Greverie C, Groenstege H, Guidi G~M, Hamdani
  S, Hayau J~F, Hebri S, Heidmann A, Heitmann H, Hello P, Hemming G, Hennes E,
  Hermel R, Heusse P, Holloway L, Huet D, Iannarelli M, Jaranowski P, Jehanno
  D, Journet L, Karkar S, Ketel T, Voet H, Kovalik J, Kowalska I, Kreckelbergh
  S, Krolak A, Lacotte J~C, Lagrange B, Penna P~L, Laval M, Marec J~C~L, Leroy
  N, Letendre N, Li T~G~F, Lieunard B, Liguori N, Lodygensky O, Lopez B,
  Lorenzini M, Loriette V, Losurdo G, Loupias M, Mackowski J~M, Maiani T,
  Majorana E, Magazz{\`{u}} C, Maksimovic I, Malvezzi V, Man N, Mancini S,
  Mansoux B, Mantovani M, Marchesoni F, Marion F, Marin P, Marque J, Martelli
  F, Masserot A, Massonnet L, Matone G, Matone L, Mazzoni M, Menzinger F,
  Michel C, Milano L, Minenkov Y, Mitra S, Mohan M, Montorio J~L, Morand R,
  Moreau F, Moreau J, Morgado N, Morgia A, Mosca S, Moscatelli V, Mours B,
  Mugnier P, Mul F~A, Naticchioni L, Neri I, Nocera F, Pacaud E, Pagliaroli G,
  Pai A, Palladino L, Palomba C, Paoletti F, Paoletti R, Paoli A, Pardi S,
  Parguez G, Parisi M, Pasqualetti A, Passaquieti R, Passuello D, Perciballi M,
  Perniola B, Persichetti G, Petit S, Pichot M, Piergiovanni F, Pietka M,
  Pignard R, Pinard L, Poggiani R, Popolizio P, Pradier T, Prato M, Prodi G~A,
  Punturo M, Puppo P, Qipiani K, Rabaste O, Rabeling D~S, R{\'{a}}cz I,
  Raffaelli F, Rapagnani P, Rapisarda S, Re V, Reboux A, Regimbau T, Reita V,
  Remilleux A, Ricci F, Ricciardi I, Richard F, Ripepe M, Robinet F, Rocchi A,
  Rolland L, Romano R, Rosi{\'{n}}ska D, Roudier P, Ruggi P, Russo G, Salconi
  L, Sannibale V, Sassolas B, Sentenac D, Solimeno S, Sottile R, Sperandio L,
  Stanga R, Sturani R, Swinkels B, Tacca M, Taddei R, Taffarello L, Tarallo M,
  Tissot S, Toncelli A, Tonelli M, Torre O, Tournefier E, Travasso F, Tremola
  C, Turri E, Vajente G, van~den Brand J~F~J, Broeck C~V~D, van~der Putten S,
  Vasuth M, Vavoulidis M, Vedovato G, Verkindt D, Vetrano F, V{\'{e}}ziant O,
  Vicer{\'{e}} A, Vinet J~Y, Vilalte S, Vitale S, Vocca H, Ward R~L, Was M,
  Yamamoto K, Yvert M, Zendri J~P and Zhang Z 2012 {\em Journal of
  Instrumentation\/} {\bf 7} P03012--P03012
  \urlprefix\url{https://doi.org/10.1088\%2F1748-0221\%2F7\%2F03\%2Fp03012}

\bibitem{EfEA2015}
Effler A, Schofield R~M~S, Frolov V~V, Gonz{\'{a}}lez G, Kawabe K, Smith J~R,
  Birch J and McCarthy R 2015 {\em Classical and Quantum Gravity\/} {\bf 32}
  035017
  \urlprefix\url{https://doi.org/10.1088\%2F0264-9381\%2F32\%2F3\%2F035017}

\bibitem{ET2011}
{ET Science Team} 2011 {\em {available from European Gravitational Observatory,
  document number ET-0106C-10}\/}

\bibitem{BBR2015}
Beker M~G, van~den Brand J~F~J and Rabeling D~S 2015 {\em Classical and Quantum
  Gravity\/} {\bf 32} 025002
  \urlprefix\url{http://stacks.iop.org/0264-9381/32/i=2/a=025002}

\bibitem{HiEA2011}
Hild S, Abernathy M, Acernese F, Amaro-Seoane P, Andersson N, Arun K, Barone F,
  Barr B, Barsuglia M, Beker M, Beveridge N, Birindelli S, Bose S, Bosi L,
  Braccini S, Bradaschia C, Bulik T, Calloni E, Cella G, Mottin E~C, Chelkowski
  S, Chincarini A, Clark J, Coccia E, Colacino C, Colas J, Cumming A,
  Cunningham L, Cuoco E, Danilishin S, Danzmann K, Salvo R~D, Dent T, Rosa R~D,
  Fiore L~D, Virgilio A~D, Doets M, Fafone V, Falferi P, Flaminio R, Franc J,
  Frasconi F, Freise A, Friedrich D, Fulda P, Gair J, Gemme G, Genin E, Gennai
  A, Giazotto A, Glampedakis K, GrÃ¤f C, Granata M, Grote H, Guidi G,
  Gurkovsky A, Hammond G, Hannam M, Harms J, Heinert D, Hendry M, Heng I,
  Hennes E, Hough J, Husa S, Huttner S, Jones G, Khalili F, Kokeyama K,
  Kokkotas K, Krishnan B, Li T~G~F, Lorenzini M, L{\"u}ck H, Majorana E, Mandel
  I, Mandic V, Mantovani M, Martin I, Michel C, Minenkov Y, Morgado N, Mosca S,
  Mours B, M{\"u}ller-Ebhardt H, Murray P, Nawrodt R, Nelson J, Oshaughnessy R,
  Ott C~D, Palomba C, Paoli A, Parguez G, Pasqualetti A, Passaquieti R,
  Passuello D, Pinard L, Plastino W, Poggiani R, Popolizio P, Prato M, Punturo
  M, Puppo P, Rabeling D, Rapagnani P, Read J, Regimbau T, Rehbein H, Reid S,
  Ricci F, Richard F, Rocchi A, Rowan S, R{\"u}diger A, SantamarÃ­a L,
  Sassolas B, Sathyaprakash B, Schnabel R, Schwarz C, Seidel P, Sintes A,
  Somiya K, Speirits F, Strain K, Strigin S, Sutton P, Tarabrin S, Th{\"u}ring
  A, van~den Brand J, van Veggel M, van~den Broeck C, Vecchio A, Veitch J,
  Vetrano F, Vicere A, Vyatchanin S, Willke B, Woan G and Yamamoto K 2011 {\em
  Classical and Quantum Gravity\/} {\bf 28} 094013
  \urlprefix\url{http://stacks.iop.org/0264-9381/28/i=9/a=094013}

\bibitem{BaHa2019}
Badaracco F and Harms J 2019 {\em Classical and Quantum Gravity\/} {\bf 36}
  145006 \urlprefix\url{https://doi.org/10.1088\%2F1361-6382\%2Fab28c1}

\bibitem{Har2019}
Harms J 2019 {\em Living Reviews in Relativity\/} {\bf 22} 6 ISSN 1433-8351
  \urlprefix\url{https://doi.org/10.1007/s41114-019-0022-2}

\bibitem{ZYL2019}
Hehua~Zhu Jinxiu~Yan W~L 2019 {\em Engineering\/} {\bf 5} 384 (pages~8)
  \urlprefix\url{http://journal.hep.com.cn/eng/EN/abstract/article\_25256.shtml}

\bibitem{ChEA2011b}
Chen G, Wu Z~z, Wang F~j and Ma Y~l 2011 {\em Environmental Earth Sciences\/}
  {\bf 62} 1667--1671 ISSN 1866-6299
  \urlprefix\url{https://doi.org/10.1007/s12665-010-0651-y}

\bibitem{MuEA2019a}
Mukund N, O'Reilly B, Somala S and Mitra S 2019 {\em Classical and Quantum
  Gravity\/} {\bf 36} 10LT01
  \urlprefix\url{https://doi.org/10.1088\%2F1361-6382\%2Fab1360}

\bibitem{MiEA2011}
Milana G, Azzara R~M, Bertrand E, Bordoni P, Cara F, Cogliano R, Cultrera G,
  Di~Giulio G, Duval A, Fodarella A, Marcucci S, Pucillo S, R{\'e}gnier J and
  Riccio G 2011 {\em Bulletin of Earthquake Engineering\/} {\bf 9} 741--759
  ISSN 1573-1456 \urlprefix\url{https://doi.org/10.1007/s10518-011-9246-6}

\bibitem{SeEA2005}
Semblat J, Kham M, Parara E, Bard P, Pitilakis K, Makra K and Raptakis D 2005
  {\em Soil Dynamics and Earthquake Engineering\/} {\bf 25} 529 -- 538 ISSN
  0267-7261 11th International Conference on Soil Dynamics and Earthquake
  Engineering (ICSDEE): Part 1
  \urlprefix\url{http://www.sciencedirect.com/science/article/pii/S0267726105000436}

\bibitem{DoEA2009}
Douglas J, Gehl P, Bonilla L~F, Scotti O, RÃ©gnier J, Duval A~M and Bertrand
  E 2009 {\em Bulletin of the Seismological Society of America\/} {\bf 99}
  1502--1520 ISSN 0037-1106 (\textit{Preprint}
  \eprint{https://pubs.geoscienceworld.org/bssa/article-pdf/99/3/1502/3670503/1502.pdf})
  \urlprefix\url{https://doi.org/10.1785/0120080075}

\bibitem{Hem2012}
Hemphill G~B 2012 {\em Practical Tunnel Construction\/} (John Wiley \& Sons,
  Ltd) ISBN 9781118350270

\bibitem{AkEA2018a}
Akutsu T, Ando M, Araki S, Araya A, Arima T, Aritomi N, Asada H, Aso Y, Atsuta
  S, Awai K, Baiotti L, Barton M~A, Chen D, Cho K, Craig K, DeSalvo R, Doi K,
  Eda K, Enomoto Y, Flaminio R, Fujibayashi S, Fujii Y, Fujimoto M~K, Fukushima
  M, Furuhata T, Hagiwara A, Haino S, Harita S, Hasegawa K, Hasegawa M, Hashino
  K, Hayama K, Hirata N, Hirose E, Ikenoue B, Inoue Y, Ioka K, Ishizaki H, Itoh
  Y, Jia D, Kagawa T, Kaji T, Kajita T, Kakizaki M, Kakuhata H, Kamiizumi M,
  Kanbara S, Kanda N, Kanemura S, Kaneyama M, Kasuya J, Kataoka Y, Kawaguchi K,
  Kawai N, Kawamura S, Kawazoe F, Kim C, Kim J, Kim J~C, Kim W, Kimura N,
  Kitaoka Y, Kobayashi K, Kojima Y, Kokeyama K, Komori K, Kotake K, Kubo K,
  Kumar R, Kume T, Kuroda K, Kuwahara Y, Lee H~K, Lee H~W, Lin C~Y, Liu Y,
  Majorana E, Mano S, Marchio M, Matsui T, Matsumoto N, Matsushima F, Michimura
  Y, Mio N, Miyakawa O, Miyake K, Miyamoto A, Miyamoto T, Miyo K, Miyoki S,
  Morii W, Morisaki S, Moriwaki Y, Muraki Y, Murakoshi M, Musha M, Nagano K,
  Nagano S, Nakamura K, Nakamura T, Nakano H, Nakano M, Nakano M, Nakao H,
  Nakao K, Narikawa T, Ni W~T, Nonomura T, Obuchi Y, Oh J~J, Oh S~H, Ohashi M,
  Ohishi N, Ohkawa M, Ohmae N, Okino K, Okutomi K, Ono K, Ono Y, Oohara K, Ota
  S, Park J, PeÃ±a~Arellano F~E, Pinto I~M, Principe M, Sago N, Saijo M,
  Saito T, Saito Y, Saitou S, Sakai K, Sakakibara Y, Sasaki Y, Sato S, Sato T,
  Sato Y, Sekiguchi T, Sekiguchi Y, Shibata M, Shiga K, Shikano Y, Shimoda T,
  Shinkai H, Shoda A, Someya N, Somiya K, Son E~J, Starecki T, Suemasa A,
  Sugimoto Y, Susa Y, Suwabe H, Suzuki T, Tachibana Y, Tagoshi H, Takada S,
  Takahashi H, Takahashi R, Takamori A, Takeda H, Tanaka H, Tanaka K, Tanaka T,
  Tatsumi D, Telada S, Tomaru T, Tsubono K, Tsuchida S, Tsukada L, Tsuzuki T,
  Uchikata N, Uchiyama T, Uehara T, Ueki S, Ueno K, Uraguchi F, Ushiba T, van
  Putten M~H~P~M, Wada S, Wakamatsu T, Yaginuma T, Yamamoto K, Yamamoto S,
  Yamamoto T, Yano K, Yokoyama J, Yokozawa T, Yoon T~H, Yuzurihara H, Zeidler
  S, Zhao Y, Zheng L, Collaboration K, Agatsuma K, Akiyama Y, Arai N, Asano M,
  Bertolini A, Fujisawa M, Goetz R, Guscott J, Hashimoto Y, Hayashida Y, Hennes
  E, Hirai K, Hirayama T, Ishitsuka H, Kato J, Khalaidovski A, Koike S, Kumeta
  A, Miener T, Morioka M, Mueller C~L, Narita T, Oda Y, Ogawa T, Sekiguchi T,
  Tamura H, Tanner D~B, Tokoku C, Toritani M, Utsuki T, Uyeshima M, van~den
  Brand J~F~J, van Heijningen J~V, Yamaguchi S and Yanagida A 2018 {\em
  Progress of Theoretical and Experimental Physics\/} {\bf 2018} ISSN 2050-3911
  013F01 (\textit{Preprint}
  \eprint{http://oup.prod.sis.lan/ptep/article-pdf/2018/1/013F01/23570266/ptx180.pdf})
  \urlprefix\url{https://doi.org/10.1093/ptep/ptx180}

\bibitem{Col2014}
Coli M and Pinzani A 2014 {\em Rock Mechanics and Rock Engineering\/} {\bf 47}
  839--851 ISSN 1434-453X
  \urlprefix\url{https://doi.org/10.1007/s00603-012-0319-x}

\bibitem{GuWa2012}
Gustafson G and Walke L 2012 {\em Hydrogeology for rock engineers\/} (BeFo
  Stockholm, Sweden)

\bibitem{Hol2014}
Holm{\o}y K~H and Nilsen B 2014 {\em Rock Mechanics and Rock Engineering\/}
  {\bf 47} 853--868 ISSN 1434-453X
  \urlprefix\url{https://doi.org/10.1007/s00603-013-0384-9}

\bibitem{Das2018}
Dassargues A 2018 {\em Hydrogeology: groundwater science and engineering\/}
  (CRC Press)

\bibitem{BMG2016}
Barakos G, Mischo H and Gutzmer J 2016 {Chapter 8 - Rare Earth Underground
  Mining Approaches with Respect to Radioactivity Control and Monitoring
  Strategies} {\em Rare Earths Industry\/} ed Lima I~B~D and Filho W~L (Boston:
  Elsevier) pp 121 -- 138 ISBN 978-0-12-802328-0
  \urlprefix\url{http://www.sciencedirect.com/science/article/pii/B9780128023280000085}

\bibitem{OjLe2014}
Ojovan M and Lee W 2014 4 - naturally occurring radionuclides {\em An
  Introduction to Nuclear Waste Immobilisation (Second Edition)\/} ed Ojovan M
  and Lee W (Oxford: Elsevier) pp 31 -- 39 second edition ed ISBN
  978-0-08-099392-8
  \urlprefix\url{http://www.sciencedirect.com/science/article/pii/B9780080993928000048}

\bibitem{YoEA2016}
Yoon J~Y, Lee J~D, Joo S~W and Kang D~R 2016 {\em Annals of occupational and
  environmental medicine\/} {\bf 28} 15--15 ISSN 2052-4374
  \urlprefix\url{https://www.ncbi.nlm.nih.gov/pubmed/27019716}

\bibitem{BMEA2004}
Burgos-Martin J, Sanchez-Padron M, Sanchez F and Martinez-Roger C 2004
  {Extremely large telescopes as a motor of socio-economic development and
  implications of their construction and installation} {\em Second Backaskog
  Workshop on Extremely Large Telescopes\/} vol 5382 ed Ardeberg A~L and
  Andersen T International Society for Optics and Photonics (SPIE) pp 142 --
  147 \urlprefix\url{https://doi.org/10.1117/12.566116}

\bibitem{Gov2008}
Govender K 2008 {\em Communicating Astronomy with the Public Proceedings from
  the IAU/National Observatory of Athens/ESA/ESO Conference\/}  160--–164

\bibitem{Nos2003}
Nosengo N 2003 {\em Nature\/} {\bf 423} 675--675 ISSN 1476-4687
  \urlprefix\url{https://doi.org/10.1038/423675b}

\bibitem{Wit2019}
Witze A 2019 {\em Nature\/} {\bf 572} 292--293
  \urlprefix\url{https://doi.org/10.1038/d41586-019-02354-5}

\bibitem{KaEA2020}
Kahanamoku S, Alegado R~A, Kagawa-Viviani A, Kamelamela K~L, Kamai B, Walkowicz
  L~M, Prescod-Weinstein C, de~los Reyes M~A and Neilson H 2020 {A Native
  Hawaiian-led summary of the current impact of constructing the Thirty Meter
  Telescope on Maunakea} \urlprefix\url{https://arxiv.org/abs/2001.00970}

\bibitem{LiLi1999}
Li C and Lindblad K 1999
  \urlprefix\url{http://urn.kb.se/resolve?urn=urn:nbn:se:ltu:diva-27070}

\bibitem{Lew2009}
Lewandowski Z and Beyenal H 2009 {\em Mechanisms of Microbially Influenced
  Corrosion\/} (Berlin, Heidelberg: Springer Berlin Heidelberg) pp 35--64 ISBN
  978-3-540-69796-1
  \urlprefix\url{https://doi.org/10.1007/978-3-540-69796-1\_3}

\bibitem{TrMe2014}
Tribollet B and Meyer M 2014 2 - ac-induced corrosion of underground pipelines
  {\em Underground Pipeline Corrosion\/} ed Orazem M~E (Woodhead Publishing) pp
  35 -- 61 ISBN 978-0-85709-509-1
  \urlprefix\url{http://www.sciencedirect.com/science/article/pii/B9780857095091500023}

\bibitem{Sch2002}
Schofield R 2002 {\em LIGO Document Server\/}
  \urlprefix\url{https://dcc.ligo.org/T020104/public}

\bibitem{FHP2003}
Fiori I, Holloway L and Paoletti F 2003 {\em Virgo Document Server\/}
  \urlprefix\url{https://tds.virgo-gw.eu/?content=3\&r=1463}

\bibitem{DaEA2004}
Daw E~J, Giaime J~A, Lormand D, Lubinski M and Zweizig J 2004 {\em Classical
  and Quantum Gravity\/} {\bf 21} 2255--2273
  \urlprefix\url{https://doi.org/10.1088\%2F0264-9381\%2F21\%2F9\%2F003}

\bibitem{FiEA2009}
Fiori I, Giordano L, Hild S, Losurdo G, Marchetti E, Mayer G and Paoletti F
  2009 {\em Proceedings of the Third International Meeting on Wind Turbine
  Noise\/} \urlprefix\url{https://tds.virgo-gw.eu/?content=3\&r=6790}

\bibitem{SaEA2011}
Saccorotti G, Piccinini D, Cauchie L and Fiori I 2011 {\em Bulletin of the
  Seismological Society of America\/} {\bf 101} 568--578 ISSN 0037-1106
  (\textit{Preprint}
  \eprint{https://pubs.geoscienceworld.org/bssa/article-pdf/101/2/568/2653359/568.pdf})
  \urlprefix\url{https://doi.org/10.1785/0120100203}

\bibitem{CoEA2017}
Coughlin M, Earle P, Harms J, Biscans S, Buchanan C, Coughlin E, Donovan F, Fee
  J, Gabbard H, Guy M, Mukund N and Perry M 2017 {\em Classical and Quantum
  Gravity\/} {\bf 34} 044004
  \urlprefix\url{http://stacks.iop.org/0264-9381/34/i=4/a=044004}

\bibitem{BiEA2018}
Biscans S, Warner J, Mittleman R, Buchanan C, Coughlin M, Evans M, Gabbard H,
  Harms J, Lantz B, Mukund N, Pele A, Pezerat C, Picart P, Radkins H and
  Shaffer T 2018 {\em Classical and Quantum Gravity\/} {\bf 35} 055004
  \urlprefix\url{https://doi.org/10.1088\%2F1361-6382\%2Faaa4aa}

\bibitem{MuEA2019}
Mukund N, Coughlin M, Harms J, Biscans S, Warner J, Pele A, Thorne K, Barker D,
  Arnaud N, Donovan F, Fiori I, Gabbard H, Lantz B, Mittleman R, Radkins H and
  Swinkels B 2019 {\em Classical and Quantum Gravity\/} {\bf 36} 085005
  \urlprefix\url{https://doi.org/10.1088\%2F1361-6382\%2Fab0d2c}

\bibitem{accadia2010noise}
Accadia T, Acernese F, Antonucci F, Astone P, Ballardin G, Barone F, Barsuglia
  M, Bauer T~S, Beker M, Belletoile A {\em et~al.\/} 2010 {\em Classical and
  Quantum Gravity\/} {\bf 27} 194011

\bibitem{AcEA2010}
Acernese F, Antonucci F, Aoudia S, Arun K, Astone P, Ballardin G, Barone F,
  Barsuglia M, Bauer T, Beker M, Bigotta S, Birindelli S, Bitossi M, Bizouard
  M, Blom M, Boccara C, Bondu F, Bonelli L, Bosi L, Braccini S, Bradaschia C,
  Brillet A, Brisson V, Budzy?ski R, Bulik T, Bulten H, Buskulic D, Cagnoli G,
  Calloni E, Campagna E, Canuel B, Carbognani F, Cavalier F, Cavalieri R, Cella
  G, Cesarini E, Chassande-Mottin E, Chincarini A, Cleva F, Coccia E, Colacino
  C, Colas J, Colla A, Colombini M, Corda C, Corsi A, Coulon J~P, Cuoco E,
  D?Antonio S, Dari A, Dattilo V, Davier M, Day R, Rosa R~D, Prete M~D, Fiore
  L~D, Lieto A~D, Emilio M~D~P, Virgilio A~D, Dietz A, Drago M, Fafone V,
  Ferrante I, Fidecaro F, Fiori I, Flaminio R, Fournier J~D, Franc J, Frasca S,
  Frasconi F, Freise A, Gammaitoni L, Garufi F, Gemme G, Genin E, Gennai A,
  Giazotto A, Granata M, Greverie C, Guidi G, Heitmann H, Hello P, Hild S, Huet
  D, Jaranowski P, Kowalska I, Królak A, Penna P~L, Leroy N, Letendre N, Li T,
  Lorenzini M, Loriette V, Losurdo G, Mackowski J~M, Majorana E, Man N,
  Mantovani M, Marchesoni F, Marion F, Marque J, Martelli F, Masserot A,
  Menzinger F, Michel C, Milano L, Minenkov Y, Mohan M, Moreau J, Morgado N,
  Morgia A, Mosca S, Moscatelli V, Mours B, Neri I, Nocera F, Pagliaroli G,
  Palomba C, Paoletti F, Pardi S, Parisi M, Pasqualetti A, Passaquieti R,
  Passuello D, Persichetti G, Pichot M, Piergiovanni F, Pietka M, Pinard L,
  Poggiani R, Prato M, Prodi G, Punturo M, Puppo P, Rabaste O, Rabeling D,
  Rapagnani P, Re V, Regimbau T, Ricci F, Robinet F, Rocchi A, Rolland L,
  Romano R, Rosi?ska D, Ruggi P, Salemi F, Sassolas B, Sentenac D, Sturani R,
  Swinkels B, Toncelli A, Tonelli M, Tournefier E, Travasso F, Trummer J,
  Vajente G, van~den Brand J, van~der Putten S, Vavoulidis M, Vedovato G,
  Verkindt D, Vetrano F, Viceré A, Vinet J~Y, Vocca H, Was M and Yvert M 2010
  {\em Astroparticle Physics\/} {\bf 33} 182 -- 189 ISSN 0927-6505
  \urlprefix\url{http://www.sciencedirect.com/science/article/pii/S0927650510000253}

\bibitem{DoEA2013}
Dooley K~L, Barsotti L, Adhikari R~X, Evans M, Fricke T~T, Fritschel P, Frolov
  V, Kawabe K and Smith-Lefebvre N 2013 {\em J. Opt. Soc. Am. A\/} {\bf 30}
  2618--2626
  \urlprefix\url{http://josaa.osa.org/abstract.cfm?URI=josaa-30-12-2618}

\bibitem{Mar2015}
Martynov D~V 2015 {\em {Lock acquisition and sensitivity analysis of advanced
  LIGO interferometers}\/} Ph.D. thesis California Institute of Technology

\bibitem{MLMa2019}
Mow-Lowry C~M and Martynov D 2019 {\em Classical and Quantum Gravity\/} {\bf
  36} 245006 \urlprefix\url{https://doi.org/10.1088\%2F1361-6382\%2Fab4e01}

\bibitem{Sau1984}
Saulson P~R 1984 {\em Phys. Rev. D\/} {\bf 30}(4) 732--736
  \urlprefix\url{http://link.aps.org/doi/10.1103/PhysRevD.30.732}

\bibitem{AkRi2009}
Aki K and Richards P~G 2009 {\em {Quantitative Seismology, 2nd edition}\/}
  (University Science Books)

\bibitem{kim2000propagation}
Kim D~S and Lee J~S 2000 {\em Soil Dynamics and Earthquake Engineering\/} {\bf
  19} 115 -- 126 ISSN 0267-7261
  \urlprefix\url{http://www.sciencedirect.com/science/article/pii/S0267726100000026}

\bibitem{WaSa1996}
Watanabe T and Sassa K 1996 {\em International Journal of Rock Mechanics and
  Mining Sciences \& Geomechanics Abstracts\/} {\bf 33} 467 -- 477 ISSN
  0148-9062
  \urlprefix\url{http://www.sciencedirect.com/science/article/pii/0148906296000058}

\bibitem{haskell1953dispersion}
Haskell N~A 1953 {\em Bulletin of the seismological Society of America\/} {\bf
  43} 17--34

\bibitem{HaOR2011}
Harms J and {O'Reilly} B 2011 {\em Bulletin of the Seismological Society of
  America\/} {\bf 101} 1478--1487 (\textit{Preprint}
  \eprint{http://www.bssaonline.org/content/101/4/1478.full.pdf+html})
  \urlprefix\url{http://www.bssaonline.org/content/101/4/1478.abstract}

\bibitem{CoEA2018b}
Coughlin M, Harms J, Bowden D~C, Meyers P, Tsai V~C, Mandic V, Pavlis G and
  Prestegard T 2019 {\em Journal of Geophysical Research: Solid Earth\/} {\bf
  124} 2941--2956
  \urlprefix\url{https://agupubs.onlinelibrary.wiley.com/doi/abs/10.1029/2018JB016608}

\bibitem{Pet1993}
Peterson J 1993 {\em Open-file report\/} {\bf 93-322}

\bibitem{haubrich1969microseisms}
Haubrich R~A and McCamy K 1969 {\em Reviews of Geophysics\/} {\bf 7} 539--571
  \urlprefix\url{https://agupubs.onlinelibrary.wiley.com/doi/abs/10.1029/RG007i003p00539}

\bibitem{LH1950}
Longuet-Higgins M~S and Jeffreys H 1950 {\em Philosophical Transactions of the
  Royal Society of London. Series A, Mathematical and Physical Sciences\/} {\bf
  243} 1--35
  \urlprefix\url{https://royalsocietypublishing.org/doi/abs/10.1098/rsta.1950.0012}

\bibitem{koley2017s}
Koley S, Bulten H~J, Brand J~v~d, Bader M, Campman X and Beker M 2017 S-wave
  velocity model estimation using ambient seismic noise at virgo, italy {\em
  SEG Technical Program Expanded Abstracts 2017\/} (Society of Exploration
  Geophysicists) pp 2946--2950

\bibitem{BoEA2002ch2}
Bormann P, Engdahl B and Kind R 2002 {\em {New Manual of Seismological
  Observatory Practice}\/} (GFZ Potsdam) chap~2

\bibitem{HaNa1998}
Hassan W and Nagy P~B 1998 {\em The Journal of the Acoustical Society of
  America\/} {\bf 104} 3107--3110
  \urlprefix\url{http://scitation.aip.org/content/asa/journal/jasa/104/5/10.1121/1.423901}

\bibitem{DABo2012}
De~Angelis S and Bodin P 2012 {\em Bulletin of the Seismological Society of
  America\/} {\bf 102} 1255--1265
  \urlprefix\url{https://doi.org/10.1785/0120110186}

\bibitem{BrEA2005}
Braccini S, Barsotti L, Bradaschia C, Cella G, Virgilio A~D, Ferrante I,
  Fidecaro F, Fiori I, Frasconi F, Gennai A, Giazotto A, Paoletti F,
  Passaquieti R, Passuello D, Poggiani R, Campagna E, Guidi G, Losurdo G,
  Martelli F, Mazzoni M, Perniola B, Piergiovanni F, Stanga R, Vetrano F,
  VicerÃ© A, Brocco L, Frasca S, Majorana E, Pai A, Palomba C, Puppo P,
  Rapagnani P, Ricci F, Ballardin G, BarillÃ© R, Cavalieri R, Cuoco E,
  Dattilo V, Enard D, Flaminio R, Freise A, Hebri S, Holloway L, Penna P~L,
  Loupias M, Marque J, Moins C, Pasqualetti A, Ruggi P, Taddei R, Zhang Z,
  Acernese F, Avino S, Barone F, Calloni E, Rosa R~D, Fiore L~D, Eleuteri A,
  Giordano L, Milano L, Pardi S, Qipiani K, Ricciardi I, Russo G, Solimeno S,
  Babusci D, Giordano G, Amico P, Bosi L, Gammaitoni L, Marchesoni F, Punturo
  M, Travasso F, Vocca H, Boccara C, Moreau J, Loriette V, Reita V, Mackowski
  J, Morgado N, Pinard L, Remillieux A, Barsuglia M, Bizouard M, Brisson V,
  Cavalier F, Clapson A, Davier M, Hello P, Krecklbergh S, Beauville F,
  Buskulic D, Gouaty R, Grosjean D, Marion F, Masserot A, Mours B, Tournefier
  E, Tombolato D, Verkindt D, Yvert M, Aoudia S, Bondu F, Brillet A,
  Chassande-Mottin E, Cleva F, Coulon J, Dujardin B, Fournier J, Heitmann H,
  Man C, Spallicci A and Vinet J 2005 {\em Astroparticle Physics\/} {\bf 23}
  557 -- 565 ISSN 0927-6505
  \urlprefix\url{http://www.sciencedirect.com/science/article/pii/S092765050500068X}

\bibitem{PLB2009}
Prieto G~A, Lawrence J~F and Beroza G~C 2009 {\em Journal of Geophysical
  Research: Solid Earth\/} {\bf 114} (\textit{Preprint}
  \eprint{https://agupubs.onlinelibrary.wiley.com/doi/pdf/10.1029/2008JB006067})
  \urlprefix\url{https://agupubs.onlinelibrary.wiley.com/doi/abs/10.1029/2008JB006067}

\bibitem{HaEA2020}
Harms J, Bonilla E~L, Coughlin M~W, Driggers J, Dwyer S~E, McManus D~J, Ross
  M~P, Slagmolen B~J~J and Venkateswara K 2020 {\em Phys. Rev. D\/} {\bf
  101}(10) 102002
  \urlprefix\url{https://link.aps.org/doi/10.1103/PhysRevD.101.102002}

\bibitem{MaEA2018}
Mandic V, Tsai V~C, Pavlis G~L, Prestegard T, Bowden D~C, Meyers P and Caton R
  2018 {\em Seismological Research Letters\/} {\bf 89} 2420
  \urlprefix\url{http://dx.doi.org/10.1785/0220170228}

\bibitem{CaEA2013}
Canuel B, Genin E, Vajente G and Marque J 2013 {\em Opt. Express\/} {\bf 21}
  10546--10562
  \urlprefix\url{http://www.opticsexpress.org/abstract.cfm?URI=oe-21-9-10546}

\bibitem{OFW2012}
Ottaway D~J, Fritschel P and Waldman S~J 2012 {\em Opt. Express\/} {\bf 20}
  8329--8336
  \urlprefix\url{http://www.opticsexpress.org/abstract.cfm?URI=oe-20-8-8329}

\bibitem{ScatteredLightO2}
Koley S, Gonzalez~Castro J, Chiummo A, Mantovani M and Fiori I 2017 Scattered
  light noise investigation at advanced virgo (lvc meeting cern, 2017)
  \urlprefix\url{https://tds.virgo-gw.eu/ql/?c=12664}

\bibitem{Hol2004}
Holton J~R and Hakim G~J 2004 {Chapter 5 The planetary boundary layer} {\em An
  Introduction to Dynamic Meteorology\/} ({\em International Geophysics\/}
  vol~88) ed Holton J~R (Academic Press) pp 115 -- 138
  \urlprefix\url{http://www.sciencedirect.com/science/article/pii/S0074614204800391}

\bibitem{Cre2008}
Creighton T 2008 {\em Classical and Quantum Gravity\/} {\bf 25} 125011
  \urlprefix\url{https://doi.org/10.1088\%2F0264-9381\%2F25\%2F12\%2F125011}

\bibitem{FiEA2018}
Fiorucci D, Harms J, Barsuglia M, Fiori I and Paoletti F 2018 {\em Phys. Rev.
  D\/} {\bf 97}(6) 062003
  \urlprefix\url{https://link.aps.org/doi/10.1103/PhysRevD.97.062003}

\bibitem{BBB2005}
Bowman J~R, Baker G~E and Bahavar M 2005 {\em Geophysical Research Letters\/}
  {\bf 32} n/a--n/a ISSN 1944-8007 l09803
  \urlprefix\url{http://dx.doi.org/10.1029/2005GL022486}

\bibitem{Gre2015}
Green D~N 2015 {\em Geophysical Journal International\/} {\bf 201} 377--389
  ISSN 0956-540X (\textit{Preprint}
  \eprint{http://oup.prod.sis.lan/gji/article-pdf/201/1/377/17367674/ggu495.pdf})
  \urlprefix\url{https://dx.doi.org/10.1093/gji/ggu495}

\bibitem{Cre2015}
Creighton T 2015 {\em LIGO document server\/}
  \urlprefix\url{https://dcc.ligo.org/LIGO-G1500688/public}

\bibitem{Beh2005}
Behrendt A 2005 {\em Temperature Measurements with Lidar\/} (New York, NY:
  Springer New York) pp 273--305 ISBN 978-0-387-25101-1
  \urlprefix\url{https://doi.org/10.1007/0-387-25101-4\_10}

\bibitem{HaEA2015b}
Hammann E, Behrendt A, Le~Mounier F and Wulfmeyer V 2015 {\em Atmospheric
  Chemistry and Physics\/} {\bf 15} 2867--2881
  \urlprefix\url{https://www.atmos-chem-phys.net/15/2867/2015/}

\bibitem{SpEA2016}
Sp\"ath F, Behrendt A, Muppa S~K, Metzendorf S, Riede A and Wulfmeyer V 2016
  {\em Atmospheric Measurement Techniques\/} {\bf 9} 1701--1720
  \urlprefix\url{https://www.atmos-meas-tech.net/9/1701/2016/}

\bibitem{CLN2004}
Chai T, Lin C~L and Newsom R~K 2004 {\em Journal of the Atmospheric Sciences\/}
  {\bf 61} 1500--1520
  \urlprefix\url{https://doi.org/10.1175/1520-0469(2004)061<1500:ROMFSF>2.0.CO;2}

\bibitem{KoEA2017}
Kowalska-Leszczynska I, Bizouard M~A, Bulik T, Christensen N, Coughlin M,
  Go{\l}kowski M, Kubisz J, Kulak A, Mlynarczyk J, Robinet F and Rohde M 2017
  {\em Classical and Quantum Gravity\/} {\bf 34} 074002
  \urlprefix\url{https://doi.org/10.1088\%2F1361-6382\%2Faa60eb}

\bibitem{AtEA2016}
Atsuta S, Ogawa T, Yamaguchi S, Hayama K, Araya A, Kanda N, Miyakawa O, Miyoki
  S, Nishizawa A, Ono K, Saito Y, Somiya K, Uchiyama T, Uyeshima M and Yano K
  2016 {\em Journal of Physics: Conference Series\/} {\bf 716} 012020
  \urlprefix\url{https://doi.org/10.1088\%2F1742-6596\%2F716\%2F1\%2F012020}

\bibitem{TCS2013}
Thrane E, Christensen N and Schofield R~M~S 2013 {\em Phys. Rev. D\/} {\bf
  87}(12) 123009
  \urlprefix\url{https://link.aps.org/doi/10.1103/PhysRevD.87.123009}

\bibitem{CoEA2018c}
Coughlin M~W, Cirone A, Meyers P, Atsuta S, Boschi V, Chincarini A, Christensen
  N~L, De~Rosa R, Effler A, Fiori I, Go\l{}kowski M, Guidry M, Harms J, Hayama
  K, Kataoka Y, Kubisz J, Kulak A, Laxen M, Matas A, Mlynarczyk J, Ogawa T,
  Paoletti F, Salvador J, Schofield R, Somiya K and Thrane E 2018 {\em Phys.
  Rev. D\/} {\bf 97}(10) 102007
  \urlprefix\url{https://link.aps.org/doi/10.1103/PhysRevD.97.102007}

\bibitem{CoEA2016b}
Coughlin M~W, Christensen N~L, Rosa R~D, Fiori I, Go{\l}kowski M, Guidry M,
  Harms J, Kubisz J, Kulak A, Mlynarczyk J, Paoletti F and Thrane E 2016 {\em
  Classical and Quantum Gravity\/} {\bf 33} 224003
  \urlprefix\url{https://doi.org/10.1088\%2F0264-9381\%2F33\%2F22\%2F224003}

\bibitem{CiEA2018}
Cirone A, Chincarini A, Neri M, Farinon S, Gemme G, Fiori I, Paoletti F,
  Majorana E, Puppo P, Rapagnani P, Ruggi P and Swinkels B~L 2018 {\em Review
  of Scientific Instruments\/} {\bf 89} 114501 (\textit{Preprint}
  \eprint{https://doi.org/10.1063/1.5045397})
  \urlprefix\url{https://doi.org/10.1063/1.5045397}

\bibitem{CiEA2019}
Cirone A, Fiori I, Paoletti F, Perez M~M, Rodr{\'{\i}}guez A~R, Swinkels B~L,
  Vazquez A~M, Gemme G and Chincarini A 2019 {\em Classical and Quantum
  Gravity\/} {\bf 36} 225004
  \urlprefix\url{https://doi.org/10.1088\%2F1361-6382\%2Fab4974}

\bibitem{AbEA2016f}
{LIGO Scientific Collaboration} 2016 {\em Classical and Quantum Gravity\/} {\bf
  33} 134001
  \urlprefix\url{https://doi.org/10.1088\%2F0264-9381\%2F33\%2F13\%2F134001}

\bibitem{AcEA2015}
Acernese F, Agathos M, Agatsuma K, Aisa D, Allemandou N, Allocca A, Amarni J,
  Astone P, Balestri G, Ballardin G, Barone F, Baronick J~P, Barsuglia M, Basti
  A, Basti F, Bauer T~S, Bavigadda V, Bejger M, Beker M~G, Belczynski C,
  Bersanetti D, Bertolini A, Bitossi M, Bizouard M~A, Bloemen S, Blom M, Boer
  M, Bogaert G, Bondi D, Bondu F, Bonelli L, Bonnand R, Boschi V, Bosi L,
  Bouedo T, Bradaschia C, Branchesi M, Briant T, Brillet A, Brisson V, Bulik T,
  Bulten H~J, Buskulic D, Buy C, Cagnoli G, Calloni E, Campeggi C, Canuel B,
  Carbognani F, Cavalier F, Cavalieri R, Cella G, Cesarini E, Chassande-Mottin
  E, Chincarini A, Chiummo A, Chua S, Cleva F, Coccia E, Cohadon P~F, Colla A,
  Colombini M, Conte A, Coulon J~P, Cuoco E, Dalmaz A, D'Antonio S, Dattilo V,
  Davier M, Day R, Debreczeni G, Degallaix J, Del{\'e}glise S, Pozzo W~D,
  Dereli H, Rosa R~D, Fiore L~D, Lieto A~D, Virgilio A~D, Doets M, Dolique V,
  Drago M, Ducrot M, Endra G, Fafone V, Farinon S, Ferrante I, Ferrini F,
  Fidecaro F, Fiori I, Flaminio R, Fournier J~D, Franco S, Frasca S, Frasconi
  F, Gammaitoni L, Garufi F, Gaspard M, Gatto A, Gemme G, Gendre B, Genin E,
  Gennai A, Ghosh S, Giacobone L, Giazotto A, Gouaty R, Granata M, Greco G,
  Groot P, Guidi G~M, Harms J, Heidmann A, Heitmann H, Hello P, Hemming G,
  Hennes E, Hofman D, Jaranowski P, Jonker R~J~G, Kasprzack M, K{\'e}f{\'e}lian
  F, Kowalska I, Kraan M, Kr{\'o}lak A, Kutynia A, Lazzaro C, Leonardi M, Leroy
  N, Letendre N, Li T~G~F, Lieunard B, Lorenzini M, Loriette V, Losurdo G,
  Magazz{\`u} C, Majorana E, Maksimovic I, Malvezzi V, Man N, Mangano V,
  Mantovani M, Marchesoni F, Marion F, Marque J, Martelli F, Martellini L,
  Masserot A, Meacher D, Meidam J, Mezzani F, Michel C, Milano L, Minenkov Y,
  Moggi A, Mohan M, Montani M, Morgado N, Mours B, Mul F, Nagy M~F, Nardecchia
  I, Naticchioni L, Nelemans G, Neri I, Neri M, Nocera F, Pacaud E, Palomba C,
  Paoletti F, Paoli A, Pasqualetti A, Passaquieti R, Passuello D, Perciballi M,
  Petit S, Pichot M, Piergiovanni F, Pillant G, Piluso A, Pinard L, Poggiani R,
  Prijatelj M, Prodi G~A, Punturo M, Puppo P, Rabeling D~S, RÃ¡cz I,
  Rapagnani P, Razzano M, Re V, Regimbau T, Ricci F, Robinet F, Rocchi A,
  Rolland L, Romano R, Rosi{\'n}ska D, Ruggi P, Saracco E, Sassolas B, Schimmel
  F, Sentenac D, Sequino V, Shah S, Siellez K, Straniero N, Swinkels B, Tacca
  M, Tonelli M, Travasso F, Turconi M, Vajente G, van Bakel N, van Beuzekom M,
  van~den Brand J~F~J, Broeck C~V~D, van~der Sluys M~V, van Heijningen J,
  Vas{\'u}th M, Vedovato G, Veitch J, Verkindt D, Vetrano F, Vicer{\'e} A,
  Vinet J~Y, Visser G, Vocca H, Ward R, Was M, Wei L~W, Yvert M, Zadro{\'c}ny A
  and Zendri J~P 2015 {\em Classical and Quantum Gravity\/} {\bf 32} 024001
  \urlprefix\url{http://stacks.iop.org/0264-9381/32/i=2/a=024001}

\bibitem{HMS2002}
Howard A, Matari{\'{c}} M~J and Sukhatme G~S 2002 {Mobile Sensor Network
  Deployment using Potential Fields: A Distributed, Scalable Solution to the
  Area Coverage Problem} {\em Distributed Autonomous Robotic Systems 5\/} ed
  Asama H, Arai T, Fukuda T and Hasegawa T (Tokyo: Springer Japan) pp 299--308
  ISBN 978-4-431-65941-9

\bibitem{CSG2018}
Cavaglià M, Staats K and Gill T 2018 {\em Communications in Computational
  Physics\/} {\bf 25} 963--987 ISSN 1991-7120
  \urlprefix\url{http://global-sci.org/intro/article\_detail/cicp/12886.html}

\bibitem{CoEA2019}
Colgan R~E, Corley K~R, Lau Y, Bartos I, Wright J~N, Marka Z and Marka S 2019
  {Efficient Gravitational-wave Glitch Identification from Environmental Data
  Through Machine Learning} \urlprefix\url{https://arxiv.org/abs/1911.11831}

\bibitem{HaEA2010}
Harms J, Acernese F, Barone F, Bartos I, Beker M, van~den Brand J~F~J,
  Christensen N, Coughlin M, DeSalvo R, Dorsher S, Heise J, Kandhasamy S,
  Mandic V, M{\'a}rka S, Mueller G, Naticchioni L, O'Keefe T, Rabeling D~S,
  Sajeva A, Trancynger T and Wand V 2010 {\em Classical and Quantum Gravity\/}
  {\bf 27} 225011
  \urlprefix\url{http://stacks.iop.org/0264-9381/27/i=22/a=225011}

\bibitem{NaEA2014}
Naticchioni L, Perciballi M, Ricci F, Coccia E, Malvezzi V, Acernese F, Barone
  F, Giordano G, Romano R, Punturo M, Rosa R~D, Calia P and Loddo G 2014 {\em
  Classical and Quantum Gravity\/} {\bf 31} 105016
  \urlprefix\url{http://stacks.iop.org/0264-9381/31/i=10/a=105016}

\bibitem{BaEA2017}
Barnaföldi G~G, Bulik T, Cieslar M, D{\'{a}}vid E, Dobr{\'{o}}ka M, Fenyvesi
  E, Gondek-Rosinska D, Gr{\'{a}}czer Z, Hamar G, Huba G, Kis {\'{A}},
  Kov{\'{a}}cs R, Lemperger I, L{\'{e}}vai P, Moln{\'{a}}r J, Nagy D,
  Nov{\'{a}}k A, Ol{\'{a}}h L, P{\'{a}}zm{\'{a}}ndi P, Piri D, Somlai L,
  Starecki T, Suchenek M, Sur{\'{a}}nyi G, Szalai S, Varga D, Vas{\'{u}}th M,
  V{\'{a}}n P, V{\'{a}}s{\'{a}}rhelyi B, Wesztergom V and W{\'{e}}ber Z 2017
  {\em Classical and Quantum Gravity\/} {\bf 34} 114001
  \urlprefix\url{https://doi.org/10.1088\%2F1361-6382\%2Faa69e3}

\bibitem{VaEA2019}
V\'an P, Barnaf\"oldi G, Bulik T, Bir\'o T, Czell\'ar S, Cie\'slar M, Czanik C,
  D\'avid E, Debreceni E, Denys M, Fenyvesi E, Gondek-Rosi\'nska D, Gr\'aczer
  Z, Hamar G, Huba G, Kov\'acs I, Kov\'acs L, Kov\'acs R, Lemperger I, L\'evai
  P, L\"ok\"os S, Moln\'ar J, Singh N, Ol\'ah L, Starecki T, Suchenek M,
  Sur\'anyi G, Tringali M~C, Varga D, Vas\'uth M, V\'as\'arhelyi B, Wesztergom
  V, W\'eber Z, Zimbor\'as Z and Somlai L 2019 {\em The European Physical
  Journal, Special Topics\/} {\bf 228} 1693--1734
  \urlprefix\url{https://doi.org/10.1140/epjst/e2019-900153-1}

\bibitem{nakamura1989method}
Nakamura Y 1989 {\em Railway Technical Research Institute, Quarterly Reports\/}
  {\bf 30}

\bibitem{acerra2004guidelines}
Acerra C, Aguacil G, Anastasiadis A, Atakan K, Azzara R, Bard P~Y, Basili R,
  Bertrand E, Bettig B, Blarel F {\em et~al.\/} 2004 {\em European
  Commission--EVG1-CT-2000-00026 SESAME\/}

\bibitem{bensen2007processing}
Bensen G~D, Ritzwoller M~H, Barmin M~P, Levshin A~L, Lin F, Moschetti M~P,
  Shapiro N~M and Yang Y 2007 {\em Geophysical Journal International\/} {\bf
  169} 1239--1260 ISSN 0956-540X
  \urlprefix\url{https://doi.org/10.1111/j.1365-246X.2007.03374.x}

\bibitem{yang2007ambient}
Yang Y, Ritzwoller M~H, Levshin A~L and Shapiro N~M 2007 {\em Geophysical
  Journal International\/} {\bf 168} 259--274 ISSN 0956-540X
  \urlprefix\url{https://doi.org/10.1111/j.1365-246X.2006.03203.x}

\bibitem{asten1984array}
Asten M~W and Henstridge J~D 1984 {\em Geophysics\/} {\bf 49} 1828--1837 ISSN
  0016-8033 \urlprefix\url{https://doi.org/10.1190/1.1441596}

\bibitem{kimman2012characteristics}
Kimman W~P, Campman X and Trampert J 2012 {\em Bulletin of the Seismological
  Society of America\/} {\bf 102} 1388--1399 ISSN 0037-1106
  \urlprefix\url{https://doi.org/10.1785/0120110069}

\bibitem{koley2018seismic}
Koley S, Campman X, Bader M, Bulten H, Brand J, Linde F and Beker M 2018
  Seismic noise characterization at a potential site for the einstein telescope
  underground gravitational wave detector {\em 80th EAGE Conference and
  Exhibition 2018\/} vol 2018 (European Association of Geoscientists \&
  Engineers) pp 1--5 ISSN 2214-4609
  \urlprefix\url{https://www.earthdoc.org/content/papers/10.3997/2214-4609.201801302}

\bibitem{woods1973plane}
Wood J~W and Lintz P~R 1973 {\em Geophysics\/} {\bf 38} 1023--1041 ISSN
  0016-8033 \urlprefix\url{https://doi.org/10.1190/1.1440393}

\bibitem{KrVi1996}
Krim H and Viberg M 1996 {\em IEEE Signal Processing Magazine\/} {\bf 13}
  67--94 ISSN 1558-0792

\bibitem{park1999multichannel}
Park C~B, Miller R~D and Xia J 1999 {\em Geophysics\/} {\bf 64} 800--808
  \urlprefix\url{https://doi.org/10.1190/1.1444590}

\bibitem{Mon2010}
Mondol N~H 2010 {\em Seismic Exploration\/} (Berlin, Heidelberg: Springer
  Berlin Heidelberg) pp 375--402 ISBN 978-3-642-02332-3
  \urlprefix\url{https://doi.org/10.1007/978-3-642-02332-3\_17}

\bibitem{WaHe2009}
Walker K and Hedlin M 2009 A review of wind-noise reduction methodologies {\em
  Infrasound Monitoring for Atmospheric Studies\/} ed Le~Pichon A, Blanc E and
  Hauchecorne A (Springer Netherlands) pp 141--182 ISBN 978-1-4020-9507-8
  \urlprefix\url{http://dx.doi.org/10.1007/978-1-4020-9508-5\_5}

\bibitem{NoEA2014}
Noble J~M, Alberts W~K, Raspet R, Collier S~L and Coleman M~A 2014 {\em
  Proceedings of Meetings on Acoustics\/} {\bf 21} 045005 (\textit{Preprint}
  \eprint{https://asa.scitation.org/doi/pdf/10.1121/2.0000307})
  \urlprefix\url{https://asa.scitation.org/doi/abs/10.1121/2.0000307}

\bibitem{Par2004}
Parker H~W 2004 Planning and site investigation in tunneling {\em 1 Congresso
  Brasileiro de T{\'u}neis e Estructuras Subterr{\^a}neas, Semin{\'a}rio
  Internacional South American Tunneling\/}

\bibitem{LiEA2015}
Li S, Tian H, Xue Y, Su M, Qiu D, Li L and Li Z 2015 {\em Journal of Coastal
  Research\/} {\bf 73} 403--409 (\textit{Preprint}
  \eprint{https://doi.org/10.2112/SI73-071.1})
  \urlprefix\url{https://doi.org/10.2112/SI73-071.1}

\bibitem{RoEA2013}
Rostami J, Sepehrmanesh M, Gharahbagh E~A and Mojtabai N 2013 {\em Tunnelling
  and Underground Space Technology\/} {\bf 33} 22 -- 33 ISSN 0886-7798
  \urlprefix\url{http://www.sciencedirect.com/science/article/pii/S0886779812001459}

\bibitem{ToEA2014}
Tom{\'a}s R, Romero R, Mulas J, Marturi{\`a} J~J, Mallorqu{\'i} J~J,
  Lopez-Sanchez J~M, Herrera G, Guti{\'e}rrez F, Gonz{\'a}lez P~J,
  Fern{\'a}ndez J, Duque S, Concha-Dimas A, Cocksley G, Casta{\~{n}}eda C,
  Carrasco D and Blanco P 2014 {\em Environmental Earth Sciences\/} {\bf 71}
  163--181 ISSN 1866-6299
  \urlprefix\url{https://doi.org/10.1007/s12665-013-2422-z}

\bibitem{JRH2017}
Johnston G, Riddell A and Hausler G 2017 {\em {The International GNSS
  Service}\/} (Cham: Springer International Publishing) pp 967--982 ISBN
  978-3-319-42928-1
  \urlprefix\url{https://doi.org/10.1007/978-3-319-42928-1\_33}

\end{thebibliography}

\end{document}